\documentclass[journal]{IEEEtran}

\usepackage{epsfig,amsmath,booktabs,dcolumn}
\usepackage{multirow}
\usepackage{tikz}
\usepackage{url}
\usepackage{cite}

\usepackage[draft]{hyperref}
\usepackage{lipsum}
\usepackage{setspace}
\usepackage{booktabs, tabularx}
\usepackage{array, xspace, subfigure, url, xcolor, amsmath, bm, array,amsfonts, balance,multirow}

\usepackage[linesnumbered,ruled,vlined]{algorithm2e}
\usepackage[noend]{algpseudocode}

\SetCommentSty{mycommfont}

\newcommand{\sys}{AccFlow\xspace}

\usepackage{mathtools}

\usepackage[font={small,bf}]{caption}

\begin{document}

\title{\sys: Defending Against the Low-Rate TCP DoS Attack in Wireless Sensor Networks}

\author{Yuan~Cao,~\IEEEmembership{Member,~IEEE,}
        Lijuan~Han,
        Xiaojin~Zhao,~\IEEEmembership{Member,~IEEE,}
        and~Xiaofang~Pan,~\IEEEmembership{Member,~IEEE}

\thanks{This work was supported in part by the National Natural Science
Foundation of China under Grant 61601168 and 61504087, in part by
the Kongque Technology Innovation Foundation of Shenzhen under Grant
KQJSCX20170727101037551, and in part by the Fundamental Research
Foundation of Shenzhen under Grant JCYJ20160520170741660 and Grant
JCYJ20170302151209762. \emph{Corresponding author: Xiaojin Zhao.}}
\thanks{Y. Cao, L. Han and X. Pan are with the College of Information Engineering, Shenzhen University, Shenzhen 518060, China.}
\thanks{X. Zhao is with the College of Electronic Science and Technology, Shenzhen University, Shenzhen 518060, China (e-mail: eexjzhao@szu.edu.cn).}
}

\markboth{IEEE SECURITY \& PRIVACY, ~Vol.~X, No.~XX,~2018}{Yuan
\MakeLowercase{\textit{et al.}}: Bare Demo of IEEEtran.cls for
Journals}

\maketitle

\begin{abstract}
Because of the open nature of the Wireless Sensor Networks (WSN), the Denial of the Service (DoS) becomes one of the most serious threats to the stability of the resource-constrained sensor nodes. In this paper, we develop \sys which is an incrementally deployable Software-Defined Networking based protocol
that is able to serve as a countermeasure against the
low-rate TCP DoS attack. The main idea of \sys is to make the attacking flows \emph{accountable} for
the congestion by dropping their packets according to their loss rates.
The larger their loss rates, the more aggressively \sys drops their packets.
Through extensive simulations, we demonstrate
that \sys can effectively defend against the low-rate TCP DoS attack even
if attackers vary their strategies by attacking at different scales and data rates. Furthermore,
while \sys is designed to solve the low-rate TCP DoS attack, we demonstrate that
\sys can also effectively defend against general DoS attacks which do not
rely on the TCP retransmission timeout mechanism but cause denial of service to legitimate users
by consistently exhausting the network resources.
Finally, we consider the scalability of \sys and its deployment in real networks.
\end{abstract}


\section{introduction}

Wireless Sensor Network (WSN) is a powerful network of widely distributed sensing, computing, storage and communication \cite{Siddiqui2018, Zidi2018}. As WSN promises to bring immense values to our daily life, it also opens a door to many challenges. The open nature, low computation capacity and limited battery power often make WSN susceptible to the to many threats \cite{Nagar2017}. One of the emerging attack is the ``Low-rate TCP DoS Attack\rq\rq{}, in which
attackers launch DoS attack by exploiting TCP retransmission timeout
mechanism \cite{low-rate}. To launch such an attack, the attackers set up
periodic on-off ``square-wave\rq\rq{} traffic whose peak
transmission rate is large enough to exhaust the network bandwidth.
When attacked, the legitimate TCP flows experience severe packet
losses and enter retransmission timeouts. If the period of the
attacking flow is close to the retransmission timeouts, the
legitimate TCP flows will face another peak when they are trying to
recover from the timeouts. As a result, they again suffer from
severe packet losses and are forced to enter even longer
retransmission timeouts. The cycle repeats and the legitimate TCP
flows are throttled to nearly zero throughput. Compared to the
general DoS attacks in which malicious users cause denial of service
to legitimate users by sending continuous high rate flows rather
than relying on the TCP retransmission timeout mechanism,
time-averaged bandwidth usage of the low-rate TCP DoS attacking flow
is small, even much less than the total available bandwidth. This is
why we call such an attack the \emph{low-rate} TCP DoS attack.

Another interesting characteristic for the low-rate TCP DoS attacking flow
is that its periodic traffic pattern is similar to that of the legitimate TCP periodic
traffic such as the video traffic that adopts the DASH \cite{dash} standard. In spite of the similar traffic pattern,
the fundamental difference between the benign TCP periodic flow and the low-rate TCP DoS attacking flow is
that the former backs off by entering retransmission timeout when its packets are lost whereas the
latter does not.

Although the low-rate TCP DoS attack has been proposed for nearly ten years, it has not been fully addressed.
Sun et al. \cite{solution1} use signal processing (autocorrelation of the traffic) to detect
the periodic burst attack at the congested router. Whenever attacks have been detected, the router traces
back to its upstream routers to find the attack source.
Such a solution may not work if the congested router has multiple upstream routers so that
the bursty traffic it detects consists of the aggregate traffic from these upstream routers.
Therefore, it is possible that the upstream routers cannot detect the bursty attacking traffic
which stops the tracing back process.
The work by Chang et al. \cite{solution2} addresses this problem
by assigning high priorities to the packets which are destined to high loss rate TCP application ports.
However, such a defense mechanism can be breached if the attackers send large volumes of traffic to a
specific protected port to cause a high loss rate at this port. Consequently, the attackers' traffic will
be marked as high priority traffic. Furthermore, because both of the aforementioned solutions target merely on
the ideal low-rate TCP DoS attack, one alternative strategy for attackers to crack the defense could
be splitting their traffic into multiple attacking flows to trigger distributed DoS attacks.
Also they do not illustrate how their defending protocols will impact the benign periodic flows
such as the aforementioned video traffic.
In this paper, we develop the \sys (representing Accountable Flow) protocol which effectively
defends against the low-rate TCP DoS attack without causing any
performance degradation to the benign periodic flows. Furthermore, \sys
also provides a strong defense against the general DoS or DDoS attacks.

Different from previous literatures, we incorporate the concept of the
Software-Defined Networking (SDN) \cite{sdn} and flow accountability when designing \sys.
The SDN architecture, in which the centralized controller makes the decision of packet routing and forwarding
so that it can explicitly execute different policies to different flows, is
proposed to provide flexibility and novelty to configure the network. For instance, one of the
benefits for such a centralized architecture is that network operators are able to
coordinate the traffic to build low latency and congestion free
networks, especially data center networks \cite{b4} \cite{swan}.
Although not proposed for solving security problems in computer networking, the concept
of SDN provides novel ways to rethink and address such problems \cite{sdns}.
Specifically, with the centralized network architecture, the controller is capable to do online
traffic monitoring and analysis. Whenever attacking flows are detected,
it blocks them and saves the network resources for
legitimate flows. The advantages of such SDN-based defending techniques are twofold.
Firstly, it responds to attacks in realtime.
Secondly, it immediately benefits the deployed routers or Autonomous Systems (ASes)
because it does not rely on reconfiguration at other parts of the network.
In spite of the aforementioned advantages, flow-based security protocols need to be scalable
to deal with huge numbers of attacking flows, especially when we consider to deploy the protocols in
Wide Area Networks (WANs). In this paper, we propose to use flow aggregation\footnote{We use
source IP address-based flow aggression. The detailed explanation is in subsections \ref{sstfsection} and \ref{flow1}.} and virtual centralized
controller\footnote{The concept of virtual centralized controller is that we use multiple coordinated
processors to serve as the centralized controller. The details are in subsection \ref{flow2}.}
to solve the scalability problem so that our protocol can be deployed
in both Local Area Networks (LANs) and WANs.

The reason why the low-rate TCP DoS attack and other kinds of DoS attacks work well
is that whenever a congestion happens, the router
drops the packets from all flows regardless of who cause
the congestion. In other words, \emph{accountability} for the congestion
is not considered for packets dropping~\cite{flowpolice}.
Consequently, legitimate TCP flows which strictly comply with the congestion avoidance
protocol and transmit at reasonable rates are equally blamed on the congestion although
it is caused by attacking flows which
send large volumes of traffic and exhaust the network bandwidth. Therefore,
in order to effectively tackle the DoS attacks, \sys
takes into consideration the accountability for congestions when dropping packets.
Specifically, the more accountable the flows are for the congestion, the more aggressively \sys
drops their packets.
We associate each flow\rq{}s accountability for the congestion to its loss rate.
The higher its loss rate, the more accountable it is for the congestion.
The reason is that attacking flows, who are more accountable for the congestion,
are always featured with high loss rates since they have
to keep sending excessive numbers of packets so as to overflow
the network. However, it is relatively rare for the legitimate TCP flows, who are less accountable for the
congestion, to suffer from consistently high loss rates because they will reduce their
transmission rates and even enter timeouts when their packets are dropped.
Therefore, the loss rate of a flow is positively correlated to its accountability for the congestion.
\sys protects the network by dropping packets from higher loss rate flows with
higher probabilities whereas dropping packets from lower loss rate flows with lower probabilities.

Through substantial amounts of simulations on ns-3 \cite{ns3}, we demonstrate that \sys
can effectively defend against both the low-rate TCP DoS attack and general DoS attacks.
In summary, the contributions of this paper are as follows.
\begin{itemize}
\item \sys is the first SDN-based security protocol that considers flow
accountability when defending against DoS or DDoS attacks.
\item We demonstrate that \sys, which does not cause any performance
degradation to benign flows, can effectively defend against
both the low-rate TCP DoS attack and general DoS attacks
even if attackers are able to vary their strategies.
\item We use flow aggregation and virtual centralized controller to solve
the scalability problem of \sys and make it deployable in real networks.
\end{itemize}

The rest of the paper is organized as follows. In section \ref{basic}, we give a brief introduction
to the low-rate TCP DoS attack. In section \ref{overview}, we elaborate on the \sys protocol.
In section \ref{effectiveness},
we thoroughly study the effectiveness of
\sys in different simulation settings. In section \ref{other},
we consider the deployment of \sys in real networks
and its interaction with other security protocols. Finally,
we conclude in section \ref{conclude}.

\section{Low-rate TCP DoS Attack}\label{basic}
\begin{figure}[h]
\centering
\subfigure[Low-rate TCP DoS attacking flow.\label{flow}]{
        \begin{minipage}[t]{0.47\textwidth}
        \includegraphics[width=\textwidth]{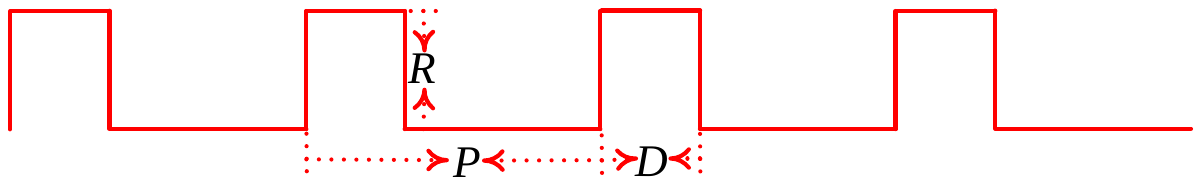}
        \end{minipage}
}
\subfigure[Network topology.\label{dumbbell}]{
        \begin{minipage}[t]{0.47\textwidth}
        \includegraphics[width=\textwidth]{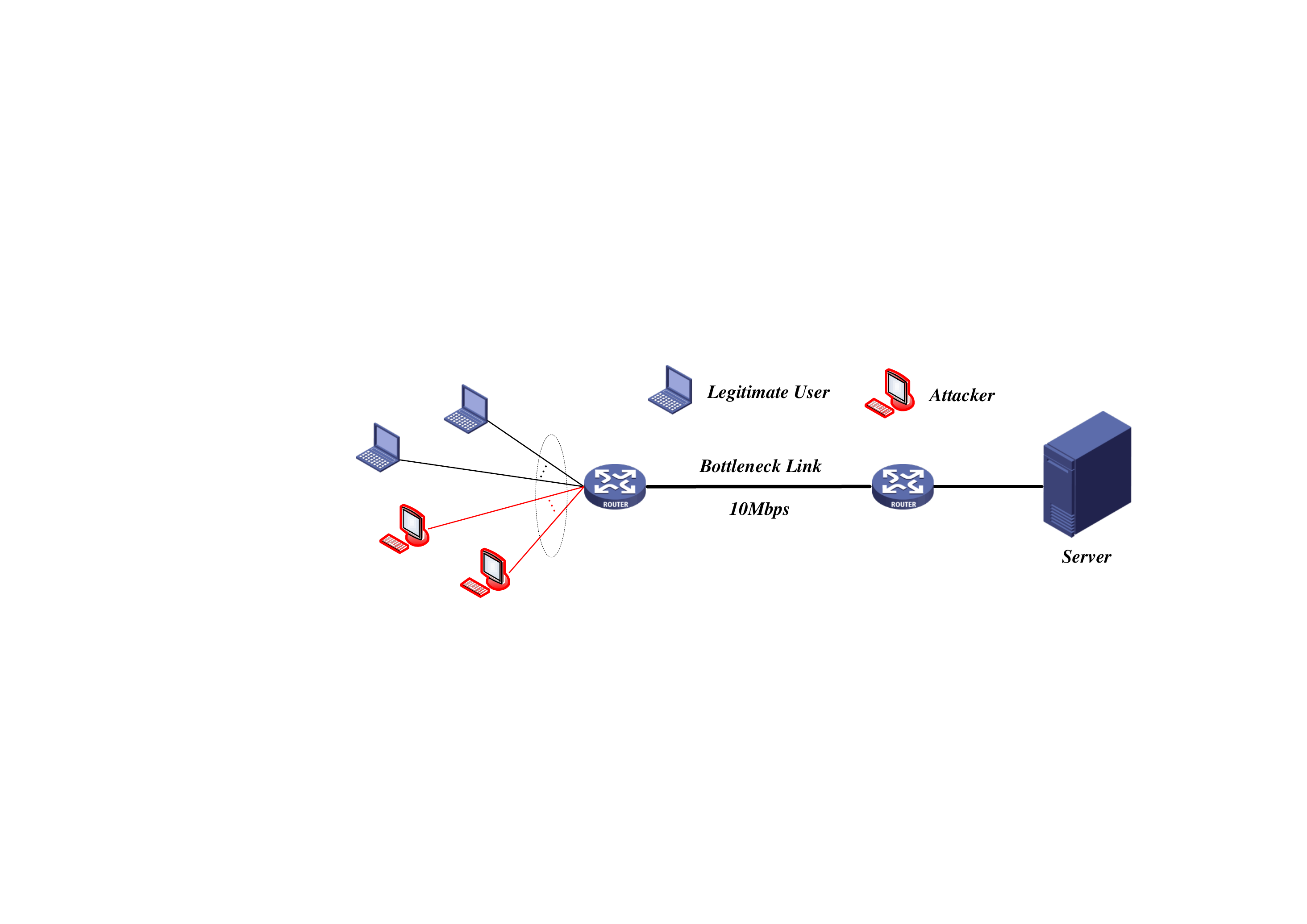}
        \end{minipage}
}
\caption{Attacking traffic and network topology.}
\end{figure}

In this section we briefly introduce the low-rate TCP DoS attack
and its effectiveness to cause denial of service to legitimate TCP flows.
The ideal low-rate TCP DoS attacking flow can be represented by a triple $\{R, P, D\}$,
where $R$ indicates the peak data rate,
$P$ indicates the attacking period and $D$ indicates the burst duration within one period, as illustrated in Figure \ref{flow}.
In order to overflow the network, $R$ needs to be larger than the bottleneck link bandwidth.
$P$ should be small enough, compared to the RTTs of
legitimate TCP flows, to attack most of the traversing flows.
$D$ is negatively correlated to $R$ if the amount of traffic generated
by the attackers in one period is fixed. A detailed discussion on the choice
of $R$, $P$ and $D$ can be found in \cite{low-rate}.

\begin{figure}[b]
\centering
\subfigure[Throughput without being attacked.\label{effectiveupper}]{
        \begin{minipage}[t]{0.45\textwidth}
        \includegraphics[width=\textwidth]{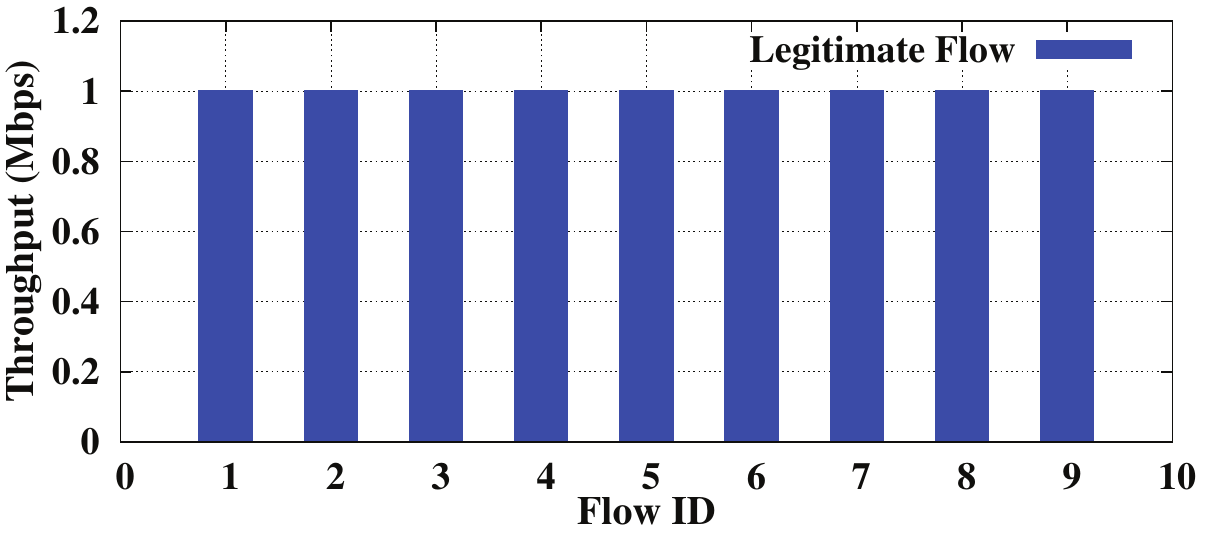}
        \end{minipage}
}
\subfigure[Throughput under attack.\label{effectivelower}]{
        \begin{minipage}[t]{0.45\textwidth}
        \includegraphics[width=\textwidth]{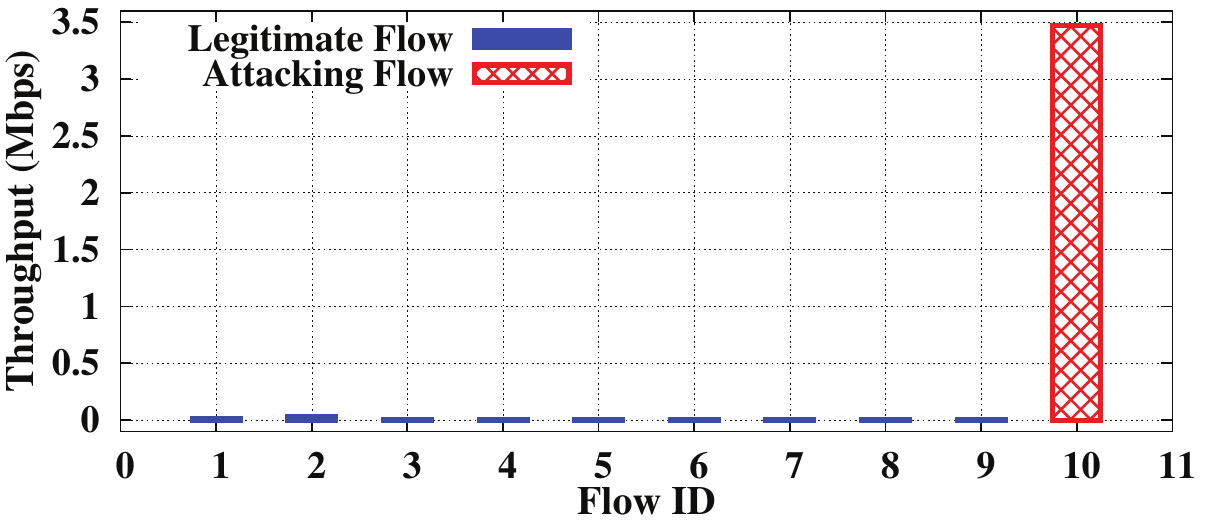}
        \end{minipage}
}
\caption{Effectiveness of the low-rate TCP DoS attack.}
\end{figure}
\begin{figure}[ht]
\begin{center}
\includegraphics[scale=0.66]{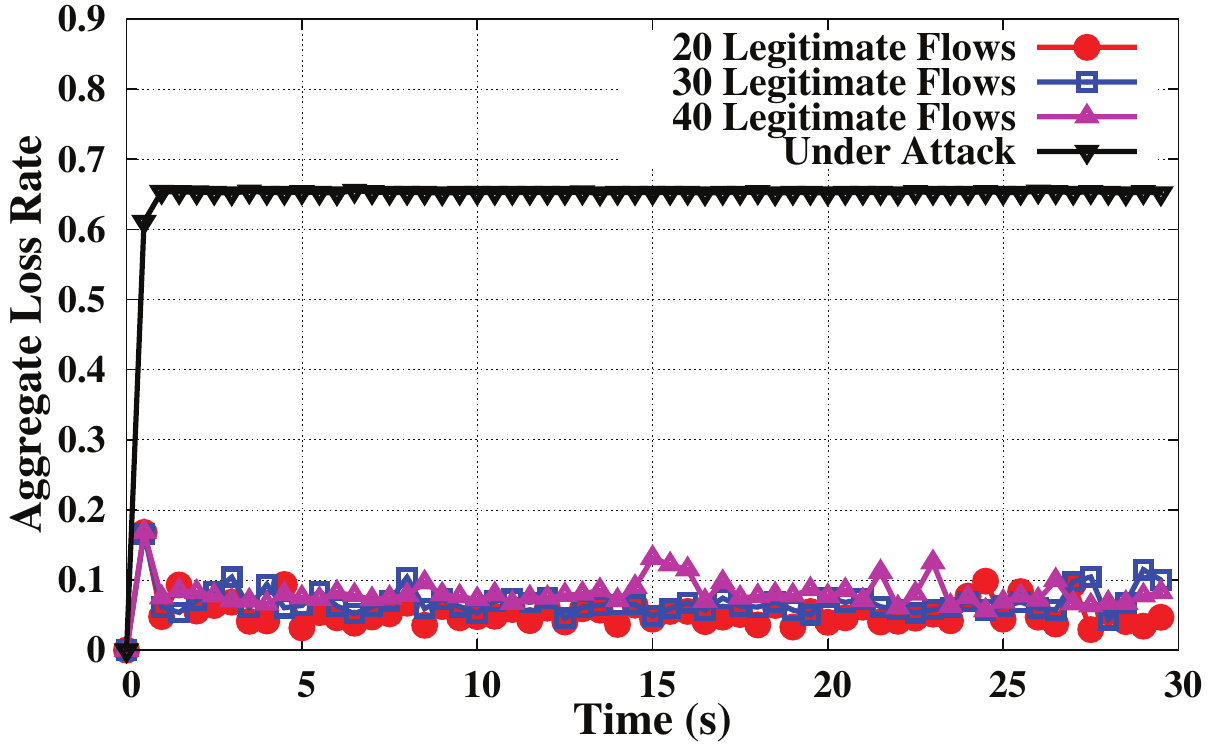}\\
\end{center}
\caption{Aggregate loss rate for normal and attacked scenarios.}\label{droprate}
\end{figure}

\begin{table*}[t]
\caption{Simulation Settings.}\label{setting}
\centering
\footnotesize
\begin{tabular}{|c|c|c|c|}
\hline
Simulation Setting&Legitimate TCP Flows&Attack Scales&Aggregate Attacking Rate\\
\hline
Setting One& 5 flows with same rate $1Mbps$ & $1$ to $50$ attacking flows  & $30Mbps$ \\
\hline
Setting Two& 9 flows with different rates ranging from $0.3Mbps$ to $1.1Mbps$ & $1$ to $50$ attacking flows & $30Mbps$ \\
\hline
Setting Three& 9 flows with different rates ranging from $0.3Mbps$ to $1.1Mbps$ & 5 attacking flows & $20Mbps$ to $60Mbps$ \\
\hline
\end{tabular}
\end{table*}

\begin{figure*}[t]
  \centering
  \mbox{
    \subfigure[\label{a}Setting One]{\includegraphics[scale=0.46]{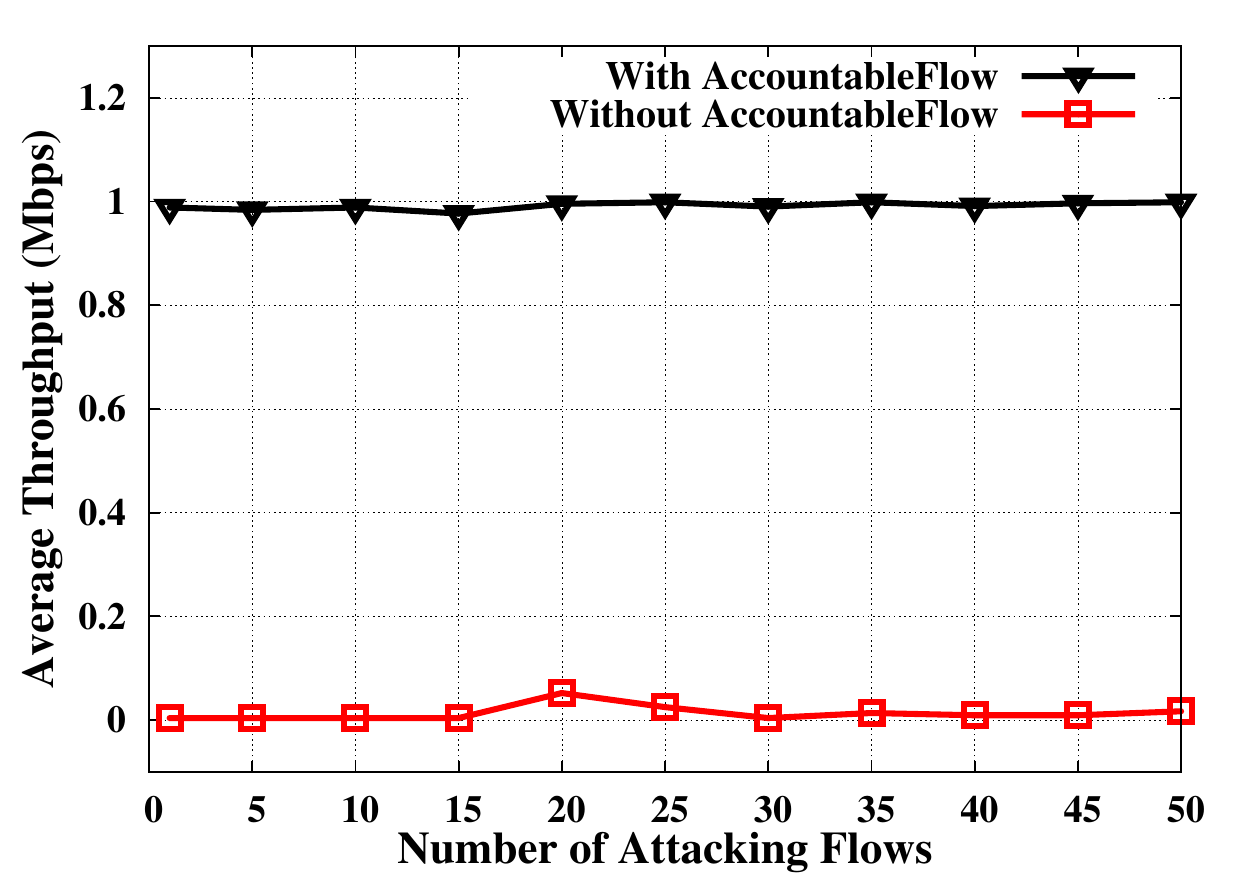}}\quad
    \subfigure[\label{b}Setting Two]{\includegraphics[scale=0.46]{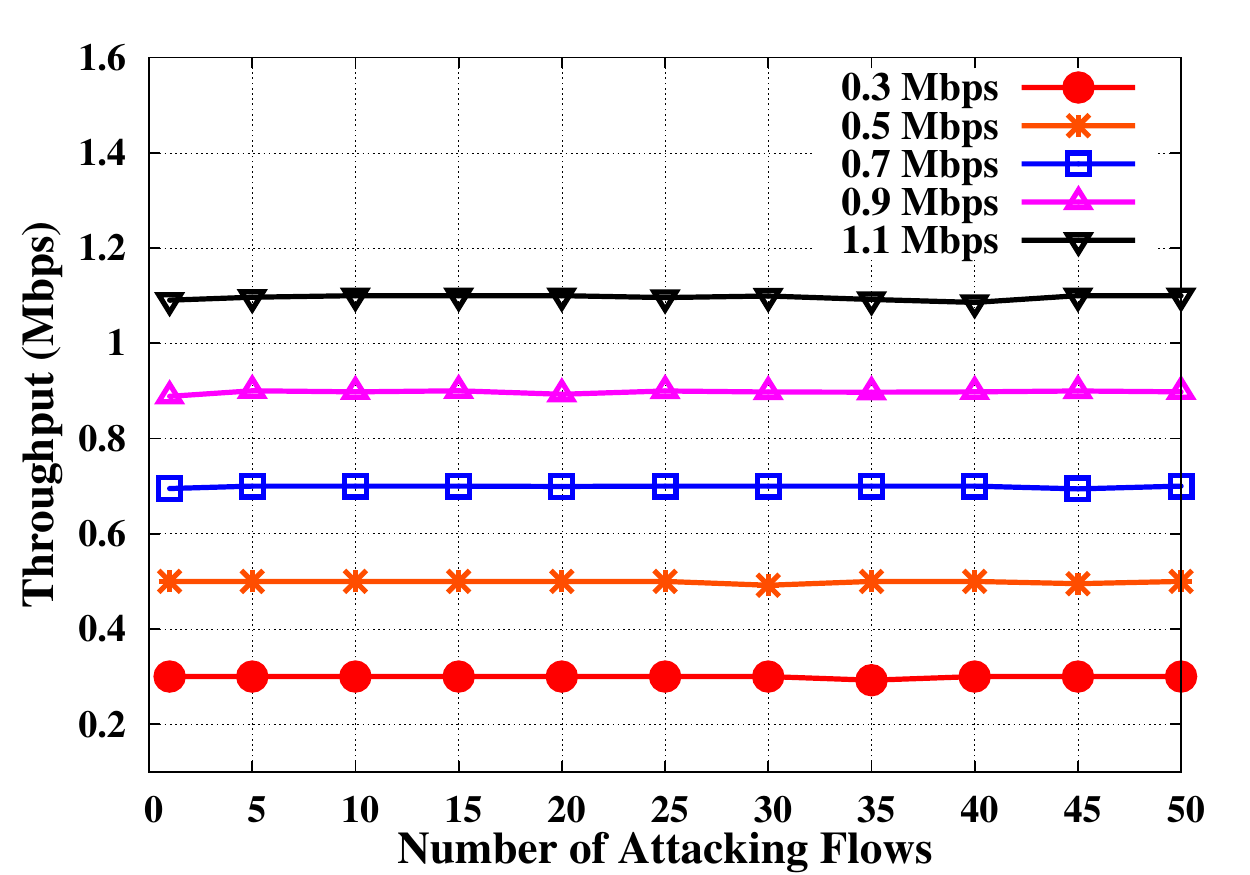}}\quad
    \subfigure[\label{c}Setting Three]{\includegraphics[scale=0.46]{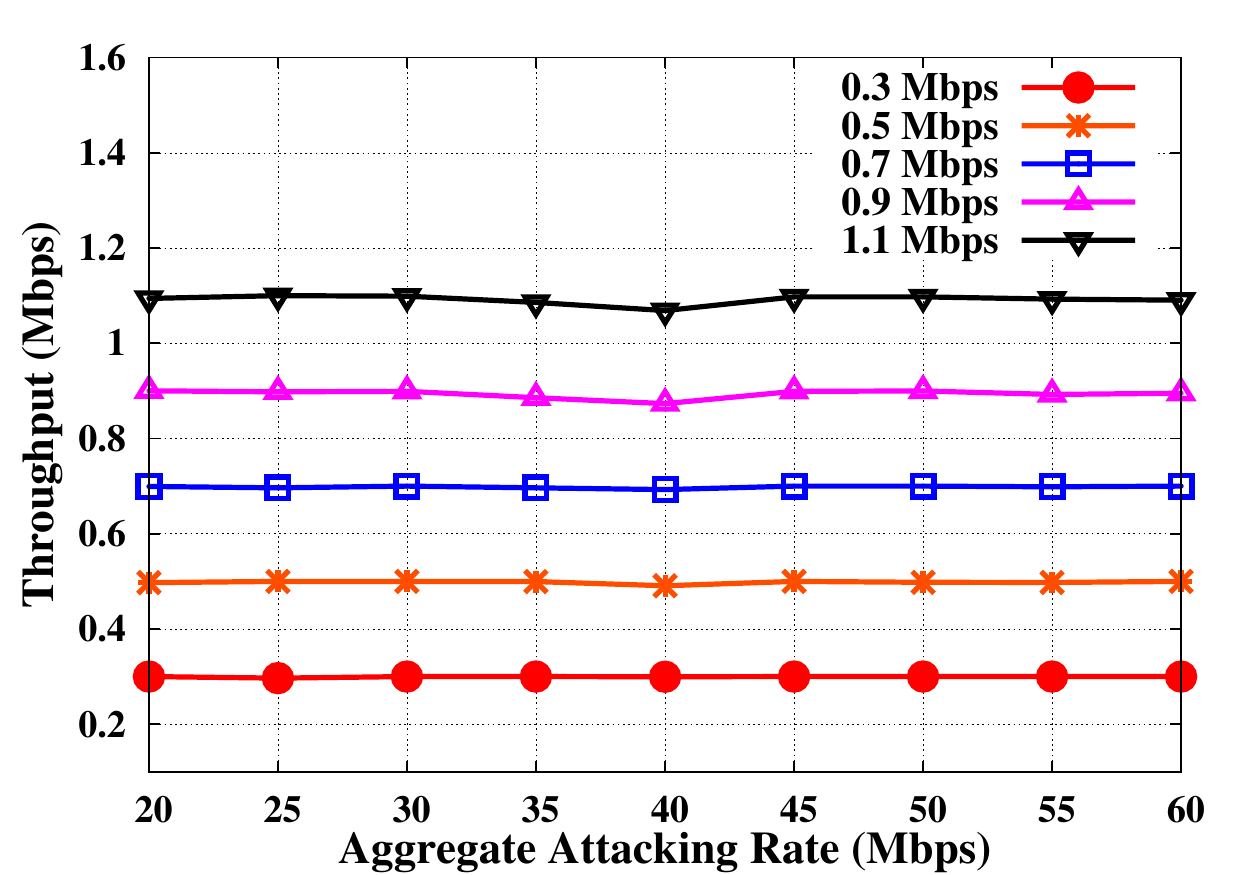}}
  }
  \caption{Under all three settings, \sys can effectively defend the legitimate TCP flows from being attacked. The achieved throughput for each
 legitimate flow is almost the same as its original transmission rate.}
  \label{earlydropeff}
\end{figure*}

\begin{figure*}[t]
  \centering
  \mbox{
    \subfigure[20 Attacking Flows in Setting One]{\includegraphics[scale=0.46]{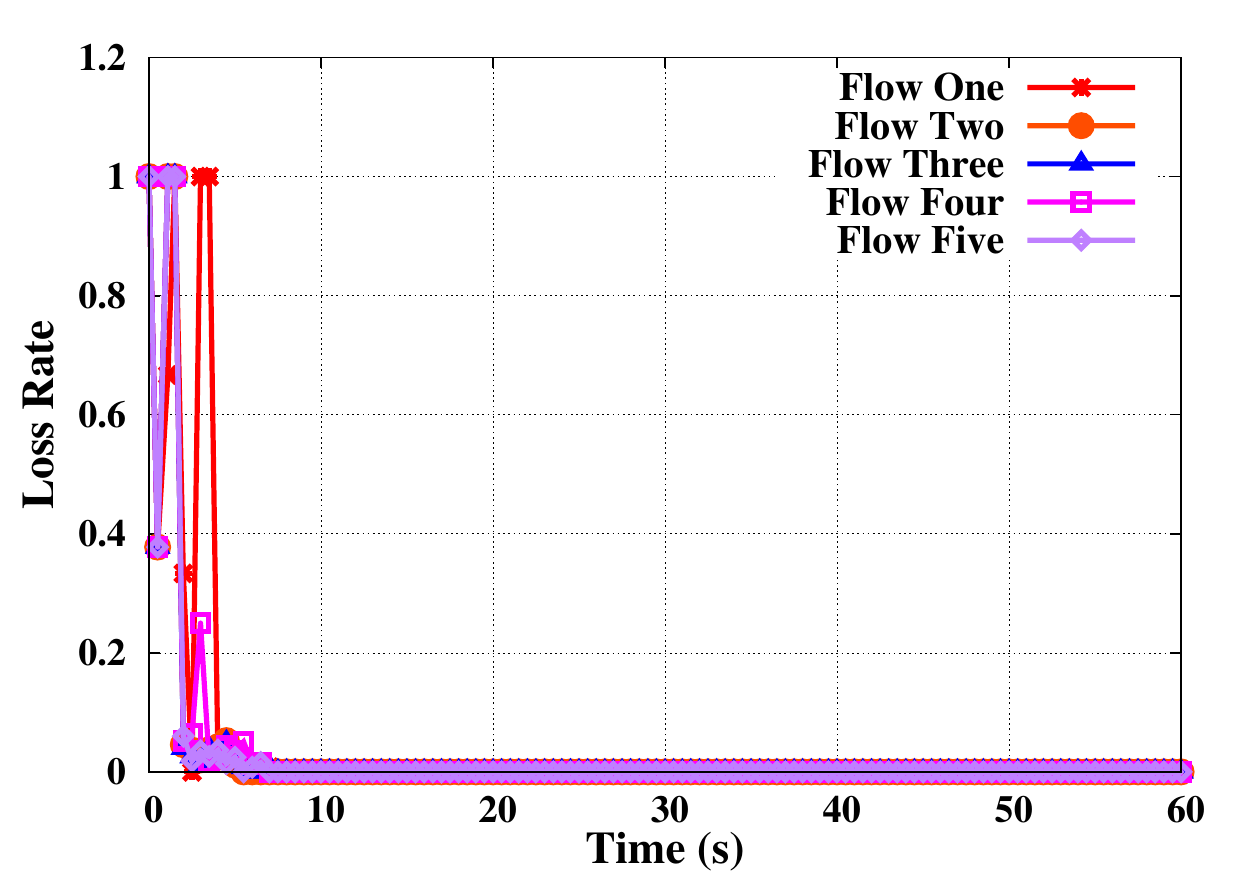}}\quad
    \subfigure[30 Attacking Flows in Setting Two]{\includegraphics[scale=0.46]{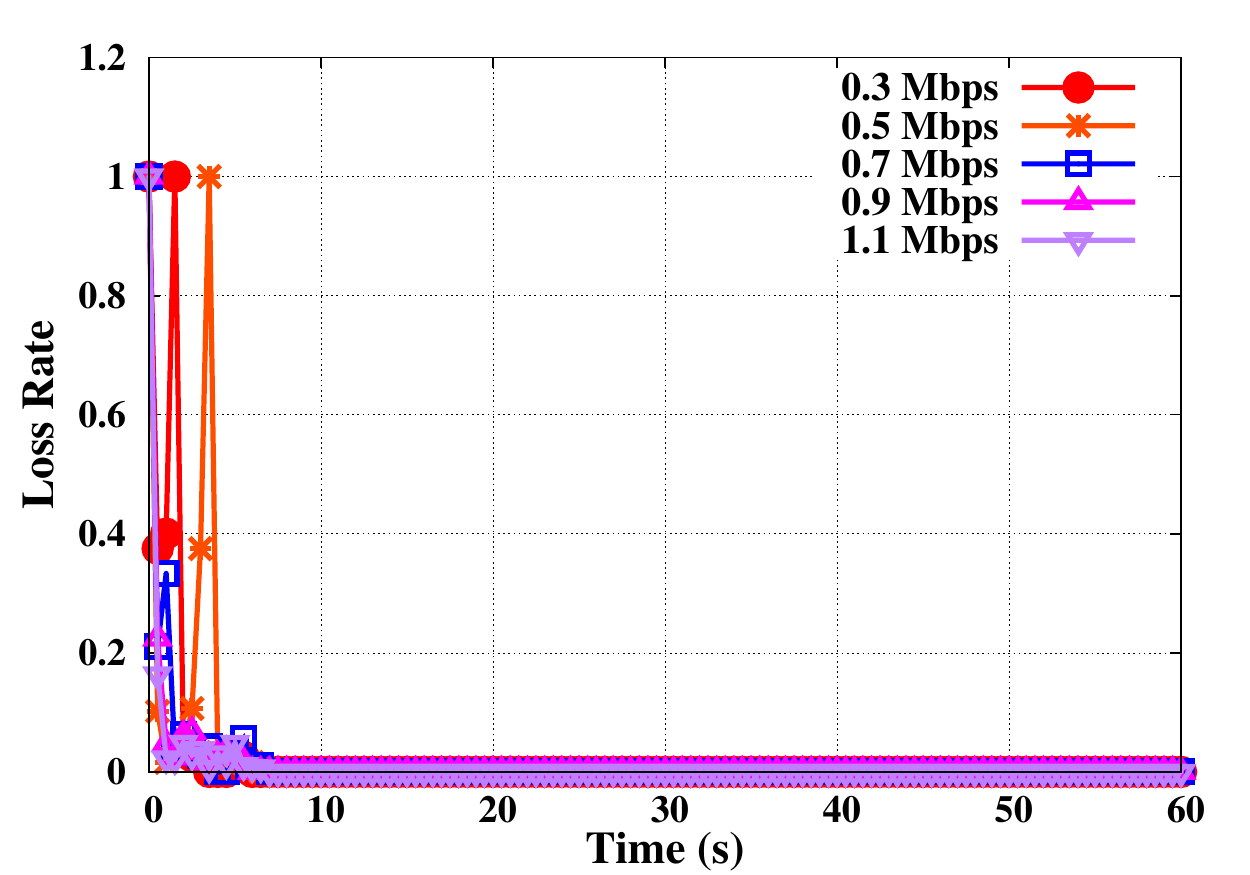}}\quad
    \subfigure[40$Mbps$ Aggregate Attacking Rate in Setting Three]{\includegraphics[scale=0.46]{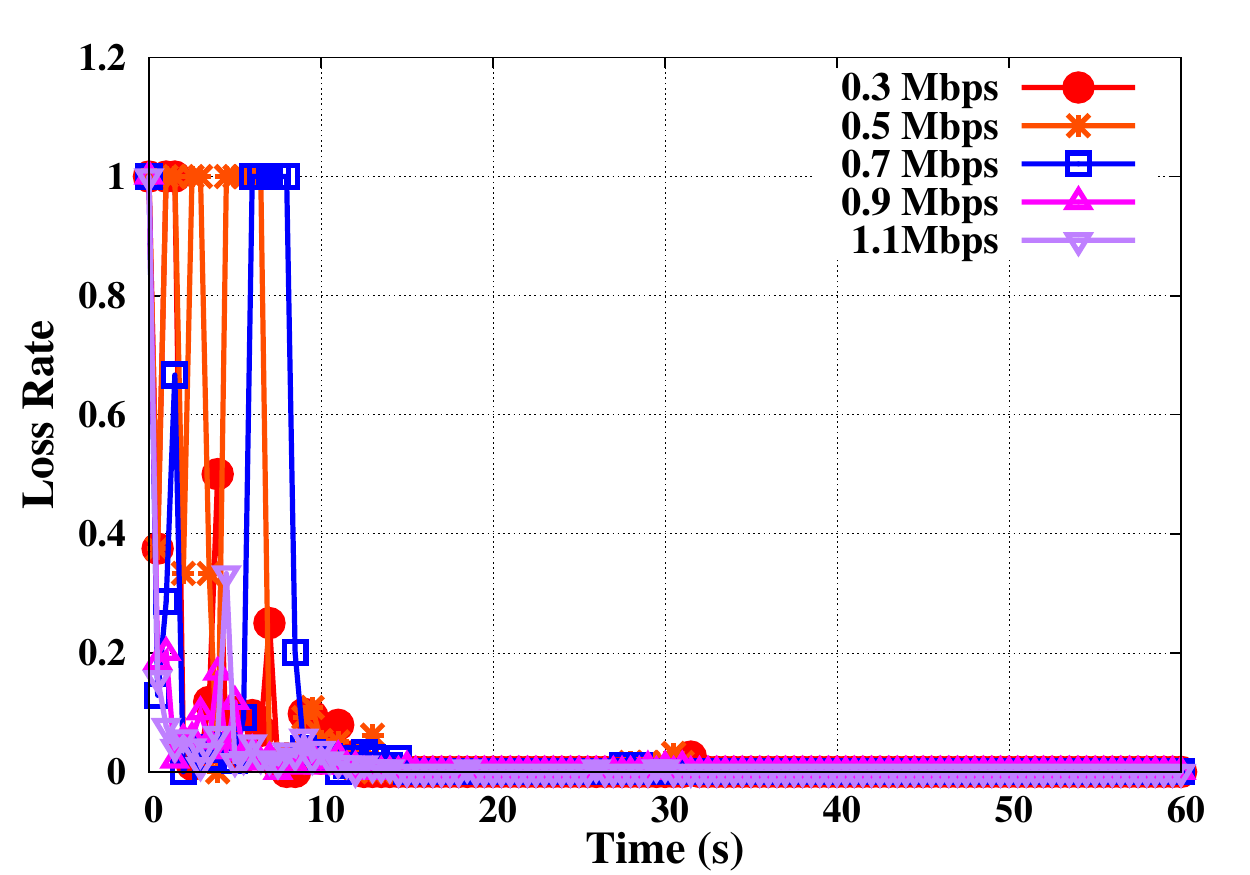}}
  }
  \caption{Convergence time of \sys.}
  \label{convergence}
\end{figure*}
We set up simulations on ns-3 platform to illustrate the effectiveness of low-rate TCP DoS attack.
We create a ``dumbbell\rq\rq{} network topology whose bottleneck link
bandwidth is $10Mbps$, as illustrated in Figure \ref{dumbbell}.
In this simulation setup, $9$ legitimate TCP flows and one attacking flow are traversing the bottleneck link.
We configure the network so that each legitimate TCP flow is transmitting at
$1Mbps$ and the attacking flow triple $\{R, P, D\}$ is
$\{30Mbps, 200ms, 67ms\}$. Note that we scale down the bottleneck link bandwidth
and flow rates in order to accelerate the simulation.
As illustrated in Figure \ref{effectiveupper}, without being attacked, all legitimate TCP flows fairly
share the bottleneck link bandwidth and achieve their desired data rates.
However, they are throttled to nearly zero throughput under attack, as illustrated Figure \ref{effectivelower}.
Note that the time-averaged data rate of the attacking flow (around $3.5Mbps$) is far less than the bottleneck link
bandwidth. Therefore, attackers use much less resources to achieve very effective DoS attacks.

\section{\sys Design}\label{overview}
\begin{figure}[h]
\begin{center}
\includegraphics[scale=0.33]{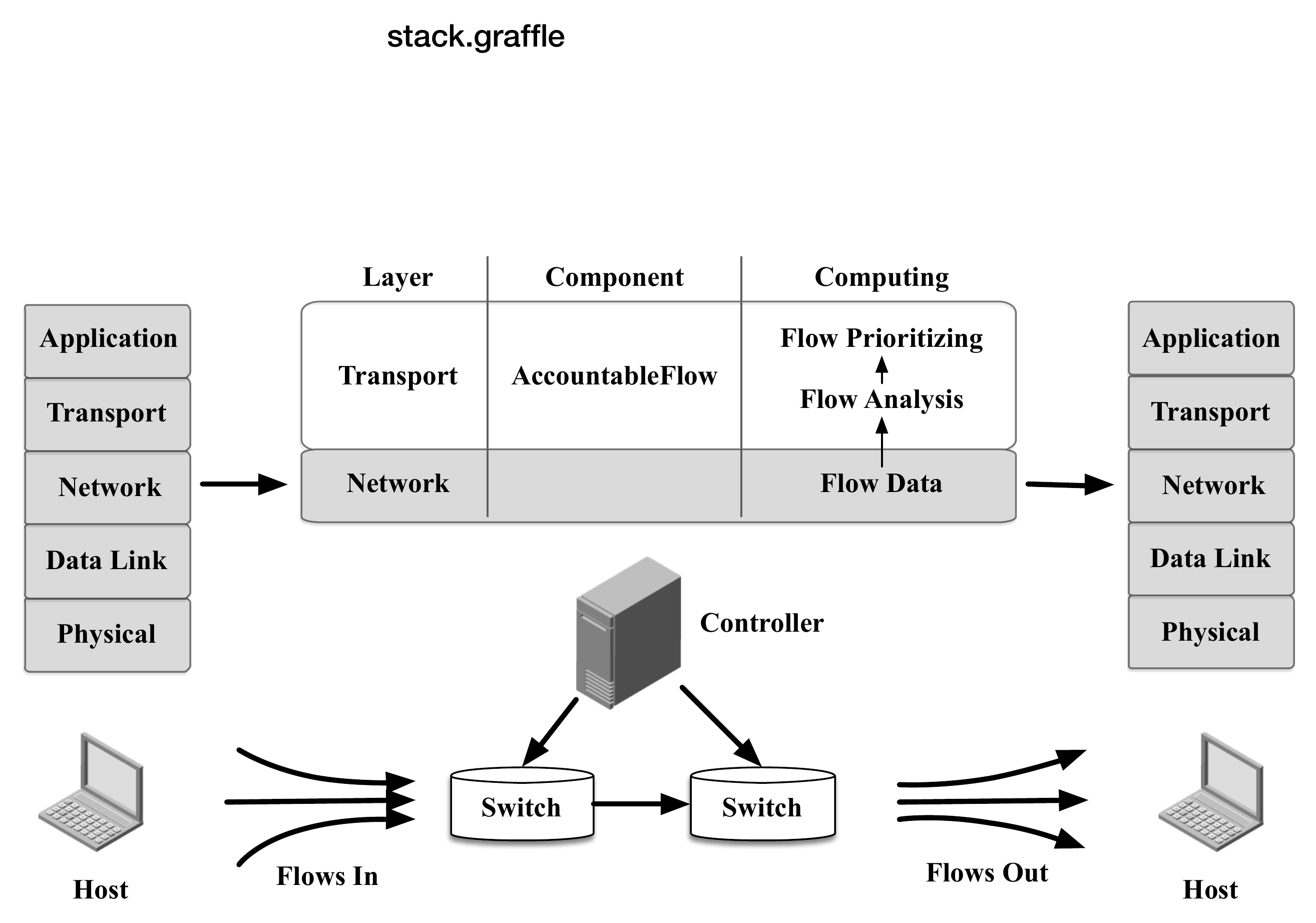}\\
\end{center}
\caption{\sys architecture.}\label{stack}
\end{figure}

In this section, we elaborate on \sys.
\sys is a transport layer protocol that is deployed on
the SDN centralized controller. The controller monitors all traversing flows,
conducts flow analysis and then instructs the switches or routers
to execute different routing and forwarding
policies to different flows. We illustrated the architecture of \sys in Figure \ref{stack}.
\sys includes two major modules, i.e., Aggressive Detection and Early Drop. Aggressive
Detection is used to block the attacking flows that are behaving aggressively
enough to be detected whereas Early Drop is used to
protect the network when attackers try to evade detection by
smartly varying their attacking strategies.

\subsection{Aggressive Detection}
\begin{figure}[h]
\centering
\subfigure[Loss rate of each flow. \label{udrupper}]{
        \begin{minipage}[t]{0.48\textwidth}
        \includegraphics[width=\textwidth]{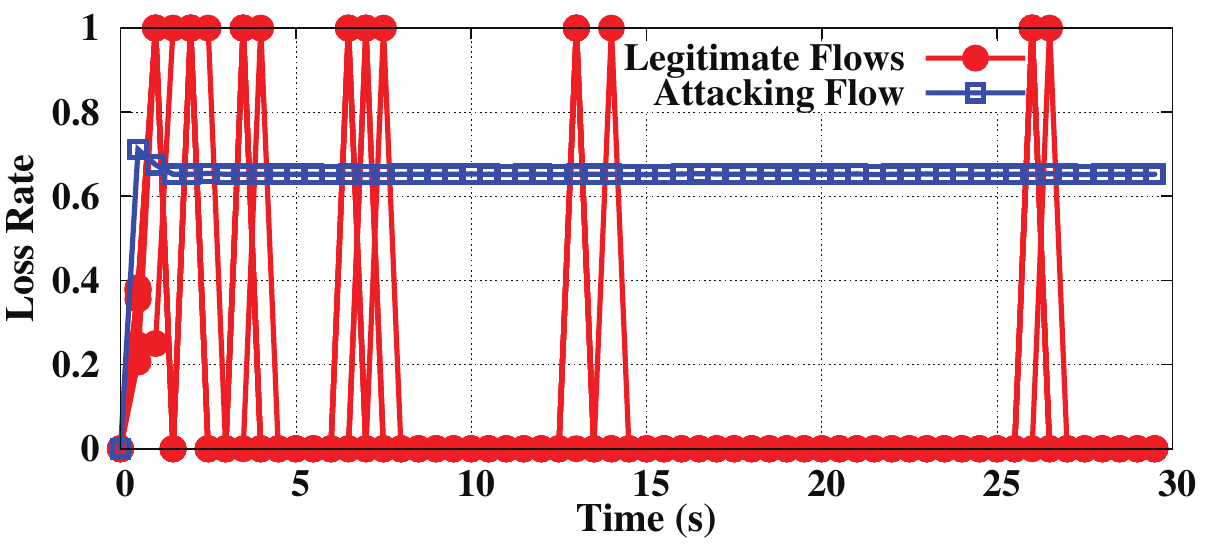}
        \end{minipage}
}
\subfigure[Uniform loss rate of each flow. \label{udrlower}]{
        \begin{minipage}[t]{0.48\textwidth}
        \includegraphics[width=\textwidth]{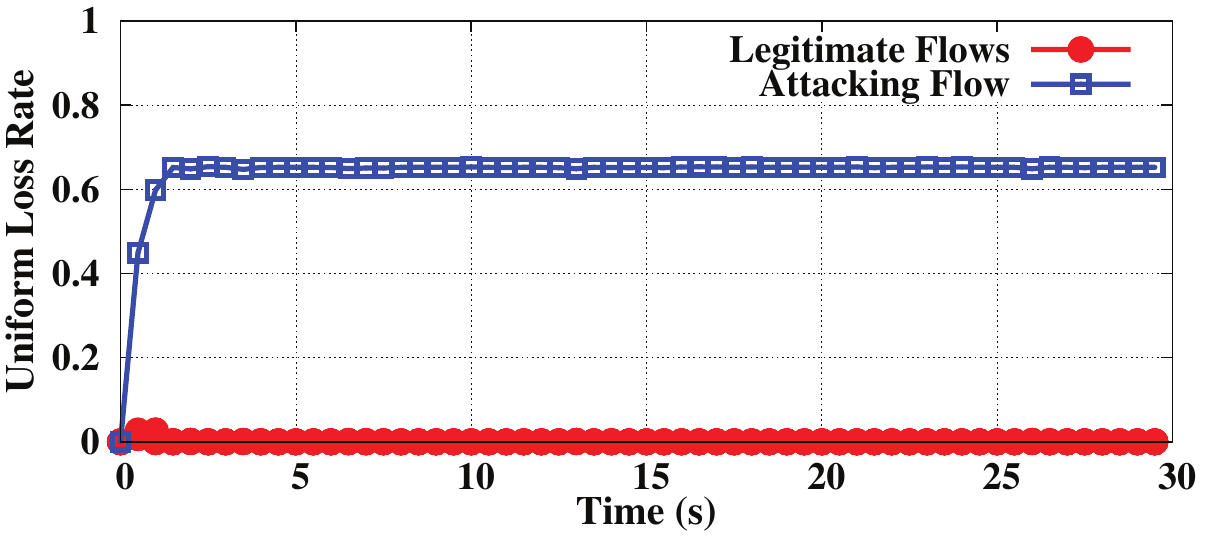}
        \end{minipage}
}
\caption{Loss rate and uniform loss rate of each flow.}
\end{figure}

In order to detect the attacking flows, we need to find unique features that distinguish them from legitimate ones.
Since the attackers consistently generate high volumes of traffic to overflow the network,
the loss rates of their flows (the ratio of lost packets over total transmitted packets)
are supposed to be higher than the legitimate flows.
Therefore, a straightforward way to differentiate legitimate flows and attacking flows is using loss rate.
To verify the effectiveness of this intuitive solution, we study the loss rate of each flow
in the simulation conducted in the previous section.
The centralized controller monitors the traffic and periodically conducts statistical
analysis for all traversing flows. The controller's flow analysis period should be
two or three times of the typical RTTs of the traversing flows so that on one
hand the controller can accurately learn the behaviors of the traversing flows and on the other hand
it can react to attacks very fast. In our simulation, we set the period to be $0.5s$, which is about
two times of the typical RTTs of the traversing flows.
In the rest of the paper, we use the term \emph{detection period} to indicate
the controller's analysis period. The simulation results are illustrated
in Figure \ref{udrupper}.

It is clear that the loss rate of the attacking flow is consistently high. However, the loss
rates for some legitimate flows in some detection periods are even higher. Therefore,
purely relying on loss rate to detect the attacking flow may result in false detection.
After analyzing the traffic sending by each flow, we realize that
the loss rate for a legitimate flow in one detection period is high because
it only sends one packet in that detection period and the only packet is dropped.
Specifically, after packet losses, the legitimate flow backs off and waits for
one retransmission timeout before entering the TCP slow start process. At the beginning of slow start,
it sends out one packet to probe the available bandwidth. If the network is extremely congested,
it is highly possible that the newly generated packet is dropped. If so, the legitimate flow is forced
to enter an even longer retransmission timeout.
This explains why the loss rate of a legitimate flow is either $1$
(the only packet is dropped) or $0$ (waiting in retransmission timeout). As for the attacking flow,
it consistently sends out huge numbers of packets and never backs off even though
a lot of the precedent packets were dropped.
Therefore, the attacking flow has a consistently high loss rate.

In our framework, we propose to use \emph{Uniform Loss Rate (ULR)} which is the product of
the loss rate and usage rate to differentiate the attacking flow from legitimate flows.
The usage rate of one flow is the ratio of the number of its transmitted packets in one
detection period over the total number of arriving packets from all flows in this detection period.
The attacking flow is featured with high ULR since it has to consistently send
large numbers of packets (high usage rate) and
never backs off even if its packets are dropped (high loss rate).
As illustrated in Figure \ref{udrlower}, there is a notable gap between
the ULR of the attacking flow and legitimate flows.
Therefore, ULR is an effective feature to leverage to differentiate the attacking flow from
legitimate ones. Whenever detecting a flow with excessively high ULR whereas other flows\rq{} ULRs are close to zero,
the centralized controller will identify it as an attacking flow and completely blocks its traffic.

\begin{figure}[h]
\centering
\subfigure[10 attacking flows. \label{ddosupper}]{
        \begin{minipage}[t]{0.47\textwidth}
        \includegraphics[width=\textwidth]{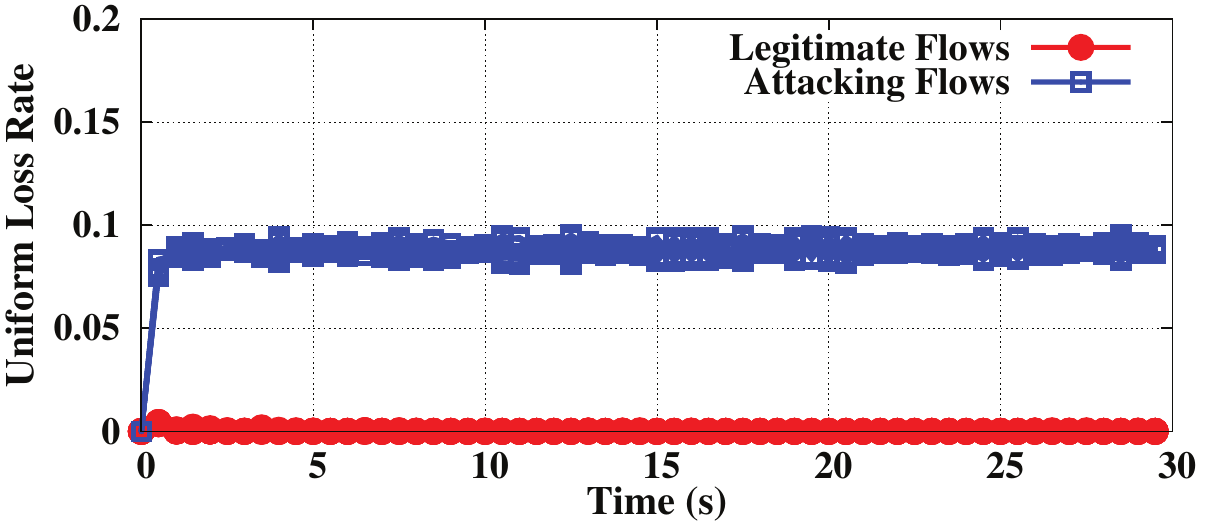}
        \end{minipage}
}
\subfigure[50 attacking flows. \label{ddoslower}]{
        \begin{minipage}[t]{0.47\textwidth}
        \includegraphics[width=\textwidth]{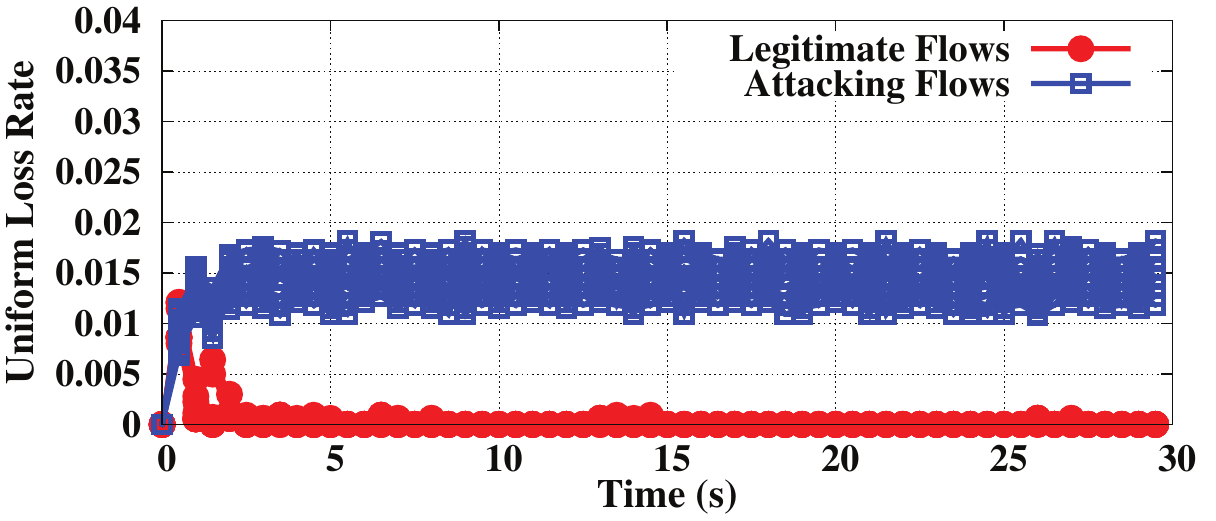}
        \end{minipage}
}
\caption{ULR of each flow under distributed DoS attack.}\label{ddos}
\end{figure}

Although simple and accurate when defending against the ideal low-rate TCP DoS attack,
Aggressive Detection requires a large enough ULR gap to differentiate between the attacking flow
and legitimate flows. Finding such a reasonable ULR threshold is difficult, especially
when attackers vary their strategies to reduce the ULRs of their flows. For example,
instead of launching one attacking flow, attackers can split their traffic into $N$ flows
and synchronize them to create the periodic burst flow. The usage rate of each individual
attacking flow will reduce by $\frac{N-1}{N}$ percentage, so does its ULR.
As the number of synchronized attacking flows increases, the ULR gaps between legitimate flows and attacking flows
decrease, which makes it difficult for the controller to detect the attacking flows.
However, since the network still experiences the same amount of attacking traffic, the DoS attack will continue to be effective.
We present the shortcoming of Aggressive Detection in Figure \ref{ddos}.
Note that when the attackers split their traffic into
$50$ synchronized flows, the ULR differences between attacking flows and legitimate flows are close to zero.
To tackle such distributed attacks, we design the Early Drop module in the next subsection.

\subsection{Early Drop}
Early Drop is proposed to effectively deal with the aforementioned distributed attacks.
The design of Early Drop is also based on the consistent flow monitoring
and periodic flow analysis by the centralized controller.
However, Early Drop does not purely rely on the ULR to detect attacking flows.
In fact, Early Drop is a heuristic algorithm illustrated in Algorithm \ref{algo}
that conducts flow-based packet dropping according to each flow's loss rate
without explicitly detecting attacking flows. Next, we elaborate on the Early Drop algorithm.
\small
\begin{algorithm}
 \If{the beginning of $k$th detection period}{
 Calculate the aggregate loss rate $\mathcal{L}$\; \label{aggre}
 \For{each traversing flow $F_i$}{
 Calculate its usage $U_i$ and loss rate $L_i$\;\label{rate}
 }
 } \label{periodEnd}
 \If{$\mathcal{L} > Th_1$}{ \label{dropLineStart}
 \For{each arriving packet}{
 Find the flow $F_j$ it belongs to;\\
 \If{$U_j> Th_2$}{ \label{accountLine}
 \If{$L_j > 0.5\times\mathcal{L}$}{ \label{reasonLineStart}
 Drop the packet with probability $L_j$\;
  \ElseIf{$N_{packet} >Th_3$}{Drop the packet with probability $L_j$\;}}\label{reasonLineEnd}
 }}}
 \caption{Early Drop}\label{algo}
\end{algorithm}
\normalsize

The algorithm is executed in every detection period.
The first \ref{periodEnd} lines of the code are executed only once at the
beginning of each detection period whereas the rest lines of the code are executed whenever a packet arrives in this detection period.
The aggregate loss rate $\mathcal{L}$ (line \ref{aggre}) is the ratio of the number of dropped packets from all flows in
the \emph{previous} detection period to the total number of arriving packets in the \emph{previous} detection period.
Similarly, usage $U_i$ of flow $F_i$ (line \ref{rate}) is the number of packets sent by $F_i$ in the previous detection period.
Loss rate $L_i$ of $F_i$ (line \ref{rate}) is the ratio of the number of dropped packets from $F_i$ in the previous detection period to $U_i$.
Note that all these values $\mathcal{L}$, $U_i$ and $L_i$ used in the current
detection period are calculated based on the statistics obtained in the previous detection period.
If the current detection period is the first detection period, the controller initializes all these values to be zero.
All these values remain the same in this detection period and will be updated at the beginning of the next detection period.
The rest lines of the codes, from line \ref{dropLineStart} to the end, conduct the packet dropping policy according to these values.

We add a condition $\mathcal{L} > Th_1$ in line \ref{dropLineStart}
for packet dropping. This is because Early Drop starts dropping packets even before the network bandwidth is exhausted.
Therefore, it is necessary to make sure that the network is being attacked before applying
such an aggressive dropping policy. In other words, Early Drop is a self
protective mechanism which automatically begins dropping packets when the network shows a sign of being attacked.
We use aggregate loss rate to verify whether the network is being attacked or not for the following reasons.
Legitimate TCP flows comply with the congestion avoidance protocol and back off when
their packets are lost due to severe congestion. Thus, even though they have large
original transmission rates, they will tailor their actual data rates to suit the available bandwidth.
Therefore, it is rare for the network to witness a consistently high aggregate loss rate under normal
situations. However, when attackers are trying to cause denial of service to
legitimate flows by exhausting the network bandwidth, they have to
continuously generate traffic even though many of their packets are dropped, i.e., they never back off
when supposed to do so. As a result, the network will experience a very high aggregate loss rate under attack.
We verify our analysis by studying the aggregate loss rate in both normal and attacked scenarios.
The experimental results, illustrated in Figure \ref{droprate}, show that even
$40$ legitimate flows each with original $1Mbps$ transmission rate are traversing the $10Mbps$ bottleneck link,
the aggregate loss rate is well bellow 10 percentage.\footnote{The real network is always bandwidth over-provisioned to
tolerate the traffic burst caused by legitimate flows, which makes it rare for the network to have a large aggregate loss rate under normal situation.}
However, when the attackers are trying to launch attack, the
aggregate loss rate is more than 65\%.
Thus, aggregate loss rate is an effective feature to indicate whether
the network is under DoS attack or not.
The network operators can have different configurations for the threshold $Th_1$ based
on their own policies and traffic characteristics. For instance,
if they want to aggressively protect their network, they need to
set a relatively low threshold for $Th_1$ and vice versa.

The main idea of the Early Drop algorithm is that it drops the packets
from \emph{accountable} flows with \emph{reasonable} probabilities.
By ``accountable'', we mean that Early Drop only blames the flows who are accountable for the congestion.
By ``reasonable'', we mean that Early Drop drops the packets of accountable flows according to their loss rates.
We achieve accountability by line \ref{accountLine} of the algorithm.
In particular, if a flow only sends one or two packets during one
detection period, it is not accountable for the congestion so that Early Drop will not drop its packets.
We add the threshold $Th_2$ to make sure that
Early Drop will not falsely blame the legitimate TCP flows who have just recovered
from retransmission timeouts and send one packet in the beginning of the TCP slow start process to probe the available bandwidth.
$Th_2$ should be small and increases with the duration of the detection period since a longer
detection period may contain more TCP slow start processes. We set $Th_2$ as 5 in our simulations
when we test the effectiveness of \sys in the next section.
Furthermore, we accomplish reasonability by considering its loss rate while
dropping packets from a particular flow (lines \ref{reasonLineStart} to \ref{reasonLineEnd}).
Specifically, Early drop divides all flows into two groups, i.e., high loss rate group and low loss rate group. All flows whose loss rates
are above half of the aggregate loss rate $\mathcal{L}$ will be categorized into the
high loss rate group and Early drop immediately drops their packets
according to their loss rates. On the contrary, flows whose loss rates are no greater than half of the $\mathcal{L}$ will
be assigned to the low loss rate group and Early Drop applies packet dropping to these flows only when the number of queueing packets
$N_{packet}$ in the router is larger than $Th_3$. The threshold $Th_3$ is used to indicate that the network
is slightly congested so that it is positively related to the router's buffer size.
We set $Th_3$ to be $10$\% of the router's buffer size in our simulations. Again
the network operators can have different configurations for $Th_3$ according to their policies.
To sum up, Early Drop blames the flows that are accountable for the congestion and the
higher their loss rates, the more aggressively it drops their packets.
The fundamental difference between Early Drop and other Active Queue Management disciplines such as
RED and WRED is that Early Drop selectively drops packets from more accountable flows (often the
attacking flows) \emph{early} before the router buffer is exhausted so that the packets from
less accountable flows (often the legitimate flows) can be enqueued.
However, RED and WRED simply drop all arriving packets when the buffer is full so
that the legitimate flows will suffer from denial of service.

Note that we can use Aggressive Detection as a patch to Early Drop. In particular, Early Drop is always active
to protect the network whereas Aggressive Detection will be applied to completely block the attacking flows
if they perform aggressively enough to be detected by the controller.
In the next section, we thoroughly test the effectiveness of \sys
through substantial amounts of simulations.

\section{Effectiveness of \sys}\label{effectiveness}
In this section, we thoroughly study the effectiveness of \sys on
ns-3 platform in four major simulation setups.
We use the network topology illustrated in Figure \ref{dumbbell} in all simulation setups whereas the
traversing flows are different under different setups.\footnote{\sys is not
limited to the simple dumbbell network topology. It is effective to protect both
the inter-domain and intra-domain traffic.}
The first setup regards the distributed low-rate TCP DoS attack, in which
we consider different types of legitimate traffic and different attacking strategies.
The second setup is about another DoS attack derived from the low-rate TCP DoS attack. We
call it Short Selfish TCP Flow (SSTF) attack because the attackers selfishly consume nearly the whole
network resources by generating excessive numbers of short TCP flows. The third setup is designed
to verify that \sys does not falsely drop packets from legitimate periodic flows.
Finally, we consider the general DoS attacks in the fourth setup.

\subsection{Distributed Low-Rate TCP DoS Attack}
In this setup, we design three different simulation settings.
In setting one, we have $5$ legitimate TCP flows each with $1Mbps$ transmission rate.
The attackers are able to launch different scales of distributed attacks by
splitting their traffic into different numbers of synchronized subflows, ranging from $1$ from $50$.\footnote{Note
that we scale down the number of flows in order to accelerate the simulation. As you can see in our experiment results, the
performance of \sys is not impacted by the scale of DDoS attacks.}
The aggregate attacking rate of all these synchronized attacking flows is about $30Mbps$, which
is three times of the bottleneck link bandwidth.
The attacking period $P$ is $200ms$ and attacking duration $D$ in one period is $67ms$. In setting two, we use
the same attacking traffic as that of setting one but we have $9$ legitimate TCP flows each with a
different transmission rate, ranging from $0.3Mpbs$ to $1.1Mpbs$.
The reason why we have both setting one and setting two is that TCP uses the Max-min fairness \cite{maxmin},
where the network first satisfies the flows with smaller demands (lower transmission rates)
and then evenly distributes the bandwidth to flows with larger demands if the network
resources are limited. As a result, in a congested network,
flows with smaller transmission rates can get their fair bandwidth
shares more easily than flows with higher rates.
Therefore, we need to consider both the two settings in our simulation.
In the third setting, we test the effectiveness of Accountable when attackers are varying their attacking rates from
$20Mbps$ to $60Mbps$. Without loss of generality, we assume that
attackers split their traffic into $5$ attacking flows in this setting and we use the same
legitimate TCP traffic as that of the setting two.
We summarize the three simulation settings in Table \ref{setting}.

The simulation results are illustrated Figure \ref{earlydropeff}.
Figure \ref{a} illustrates the results for setting one when five $1Mbps$ legitimate TCP flows
are traversing the network.
Since the whole bandwidth of the bottleneck link is $10Mbps$, which
is large enough to hold all the legitimate traffic, all the $5$ flows
should not experience any packet losses\footnote{In this paper, we only consider packet losses
caused by network congestions.} and achieve their ideal throughput when they are not attacked.
As we can see in Figure \ref{a}, \sys can effectively protect the legitimate flows
from being attacked since the average throughput of the legitimate flows is close to the desired transmission rate.
Furthermore, the performance of \sys does not degrade
as the number of attacking flows increases. Figure \ref{b} illustrates the simulation results for
setting two. Note that we set up $9$ legitimate flows in this setting but we only plot
the results for $5$ of them in the figure for concise presentation.
With \sys, all $9$ legitimate flows are able to achieve their desired data rates even
under large scales of attacks. In setting three, we vary the aggregate attacking rates
from $20Mbps$ to $60Mbps$.
Again we plot the simulation results for $5$ legitimate flows in the figure.
The results show that \sys can also effectively defend the
network even if attackers are able to change their attacking rates.

Here, we make a detailed explanation for the effectiveness of \sys.
Consider one legitimate TCP flow and one attacking flow in our simulations.
Assume that attackers launch DoS attacks in detection period $T_k$.
Due to severe packet losses in $T_k$, the legitimate TCP flow envisions a heavily congested
network and enters retransmission timeout. Therefore, in the next detection period $T_{k+1}$, either its loss rate is
zero if it is still waiting in the timeout or its usage is very low if
it just recovers from the timeout and sends small amounts of packets to probe the available bandwidth.
Under both scenarios, \sys will not further blame the legitimate flow. However, the attackers have to continuously send
high volumes of traffic in order to overflow the bottleneck link. Thus the attacking flow
still has a large usage in the next detection period $T_{k+1}$ in spite of its high loss rate in $T_k$.
Under such a situation, \sys will early drop its packets according to
its loss rate. Furthermore, if the network is still congested after early drop,
the router itself will also drop packets since it cannot deal with so much traffic.
Therefore, the attacking flow will experience an even larger loss rate in detection period $T_{k+1}$.
This cycle repeats so that the loss rate of the attacking flow increases in each detection period until the network is
not congested or its loss rate equals to one. Under both scenarios, the attacking flow can no longer
harm the network.

In order to be a realtime defending technique, \sys needs to react to attacks very fast.
Here we study the convergence time of \sys, i.e., how long it
takes \sys to clear up the attacks.
We define the convergence time as the time when all legitimate flows' loss
rates are zero. Without loss of generality, we randomly pick up one simulation setup in each
of the three settings listed in Table \ref{setting} to study its convergence time. Specifically, we
test the scenarios where attackers set up $20$ and $30$ attacking flows in setting one and setting two, respectively.
As for the setting three, we use the case when attackers are generating traffic at rate of $40Mbps$.
Our simulation results are illustrated in Figure \ref{convergence}.
Under all three scenarios, \sys can react to the attack quickly and the
convergence time is in the order of seconds or tens of seconds.

As a flow-based defending technique, \sys needs to be scalable to deal with large numbers of attacking flows.
Although the performance of \sys does not decline
as the number of attacking flows increases (as illustrated in our simulation results),
huge numbers of flows will exhaust the CPU and storage of the centralized controller.
We discuss the scalability problem and propose our solutions in section \ref{other} where
we consider the deployment of \sys in real networks.

\subsection{Short Selfish TCP Flow Attack}\label{sstfsection}
\begin{figure}[ht]
\begin{center}
\includegraphics[scale=0.65]{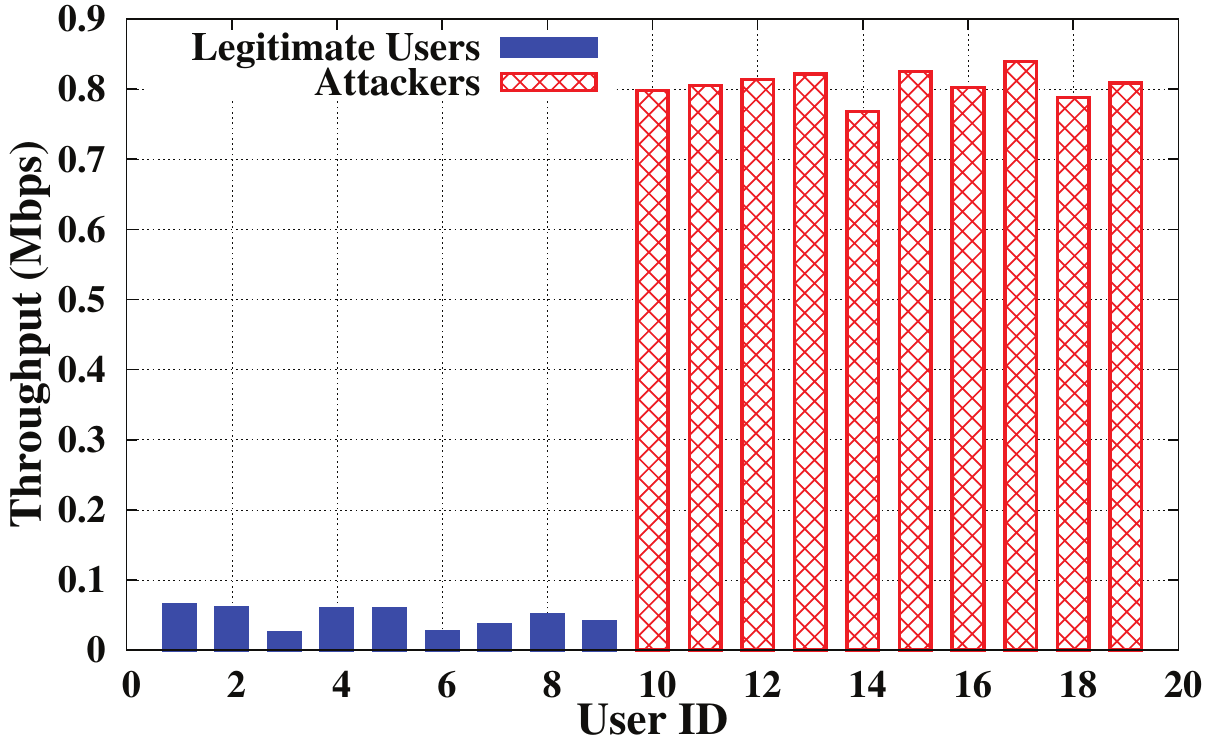}\\
\end{center}
\caption{Effectiveness of the SSTF attack.}\label{sstf}
\end{figure}

In this section we discuss a very effective DoS attack which is similar to but still fundamentally
different from the low-rate TCP DoS attack.
The attacking technique is that malicious users periodically set up many short TCP flows to gain unfair
share of network resources. Specifically,
the early coming short TCP flows congest the
network and cause all flows (including both legitimate TCP flows and themselves)
to enter retransmission timeouts. Then the attackers selfishly start new short TCP flows to occupy
the whole network bandwidth since no one besides attackers is transmitting now.
The interesting point of such DoS attacks is that it seems that the attackers
never deviate from the TCP protocol since these
short TCP flows will back off when their packets are lost.
However, the attackers are able to cause very severe denial of service to legitimate users simply by breaking their
traffic into small short flows. We name such an attack the Short Selfish TCP Flow (SSTF) attack.
Note that the difference between the SSTF attack and low-rate TCP DoS attack is that the former cannot
synchronize all these short TCP flows to create the regular periodic burst traffic since transmitters have
to wait for the ACKs before sending new packets.
We set up simulations to illustrate the effectiveness of the SSTF attack.
In our simulation we have $10$ attackers and each of them sends one
short TCP flow with $3Mbps$ data rate every $200ms$. Furthermore,
we have $9$ legitimate TCP flows with $1Mbps$ transmission rate each.
The results, illustrated in Figure \ref{sstf}, show that the SSTF attack is able to effectively
throttle the legitimate users to almost zero throughput.

Now we explain why the SSTF attack can cause such an effective denial of service to the legitimate users
even though each individual short flow itself behaves exactly the same as a legitimate TCP flow.\footnote{
Here we mean each short TCP flow complies with the TCP protocol and will back off when a congestion happens.}
First let us reconstruct the attacking procedure. Assume that attackers
first set up $n$ short TCP flows $\mathcal{A}=\{A_1, A_2, ..., A_n\}$ to cause congestion.
Then all the legitimate flows and attacking flows in $\mathcal{A}$ will
suffer from severe packet losses and enter retransmission timeouts.
After a short period, attackers set up another set of short TCP flows $\mathcal{B}=\{B_1, B_2, ..., B_m\}$.
Since no one is transmitting now, $\mathcal{B}$
will occupy the whole network resources. When the legitimate flows try to recover from timeouts, they may face another set of short attacking
flows started by the attackers after flows in $\mathcal{B}$ finish.
Again, congestion happens and the legitimate flows are forced to enter even longer retransmission timeouts.
The cycle repeats and the attackers are able to selfishly utilize nearly the whole
network resources. In a word, by sacrificing a small fraction of their traffic, the attackers create a ``clear\rq\rq{} network environment
for most of their traffic and cause severe denial of service to the legitimate users.

The trick played by the attackers is to evade accountability by continuously generating fresh short flows.
In particular, it is the flows in $\mathcal{A}$ that cause the congestion so that
we have no reason to blame the flows in $\mathcal{B}$.
However, when flows in $\mathcal{A}$ experience severe packet losses, the source should realize that the network is congested
and should not start new flows. Thus, flow set $\mathcal{B}$ should not be generated because both $\mathcal{A}$ and
$\mathcal{B}$ come from the same source (attackers). Therefore, although each individual short TCP flow complies with the
TCP protocol, the attackers still behave maliciously by periodically setting up new TCP flows even though
the previous flows experience high loss rates.

We propose to use \emph{flow aggregation} to defend against the SSTF attack.
Specifically, all flows with the same source IP address will be aggregated as
one flow.\footnote{It is possible to conduct flow aggregation based on other properties such as
source IP address and application port pair. We leave the discussion of different aggregating properties
in future works.} Consequently, although flow $A_i\in \mathcal{A}$ and flow $B_j\in\mathcal{B}$ are
different flows, they may have the same source IP address
since they are both generated by the attackers.
As a result, \sys will aggregate them as one flow. Therefore,
flow $B_j$ will be blamed for the congestion caused by flow $A_i$.
Similarly, the subsequent flows will be accountable for the congestion caused by their precedent flows
as long as they have the same source IP address.
Thus, the attackers are not able to selfishly over-utilize the network resources by
creating new flows. A potential problem for conducting such flow aggregation is that
attackers can spoof their source IP addresses to keep
generating new flows. However, the network security community has proposed effective
mechanisms such as Stackpi \cite{ip1} and packet filters \cite{ip2} to prevent source IP spoofing.
\sys can embrace such security protocols to prevent attackers from faking flows.
Another potential problem is that the Network Address Translation (NAT) router translates
the IP addresses of the hosts within its LAN to its public IP address.
Then all the flows from different hosts will have the same source IP address after they leave the LAN.
When they reach other remote sites which deploy \sys, they will be aggregated as one flow.
Thus, a single compromised host within the LAN may cause denial of service to all the legitimate hosts
within the LAN since their flows are aggregated as the same flow. To solve the problem,
we can deploy \sys on the NAT router so that it will drop the packets from the
local attacking flow and save bandwidth for legitimate flows. Therefore, the local attacking flow will not
be able to leave the LAN to attack the remote sites.

\begin{figure*}[t]
  \centering
  \mbox{
    \subfigure[Setting One]{\includegraphics[scale=0.46]{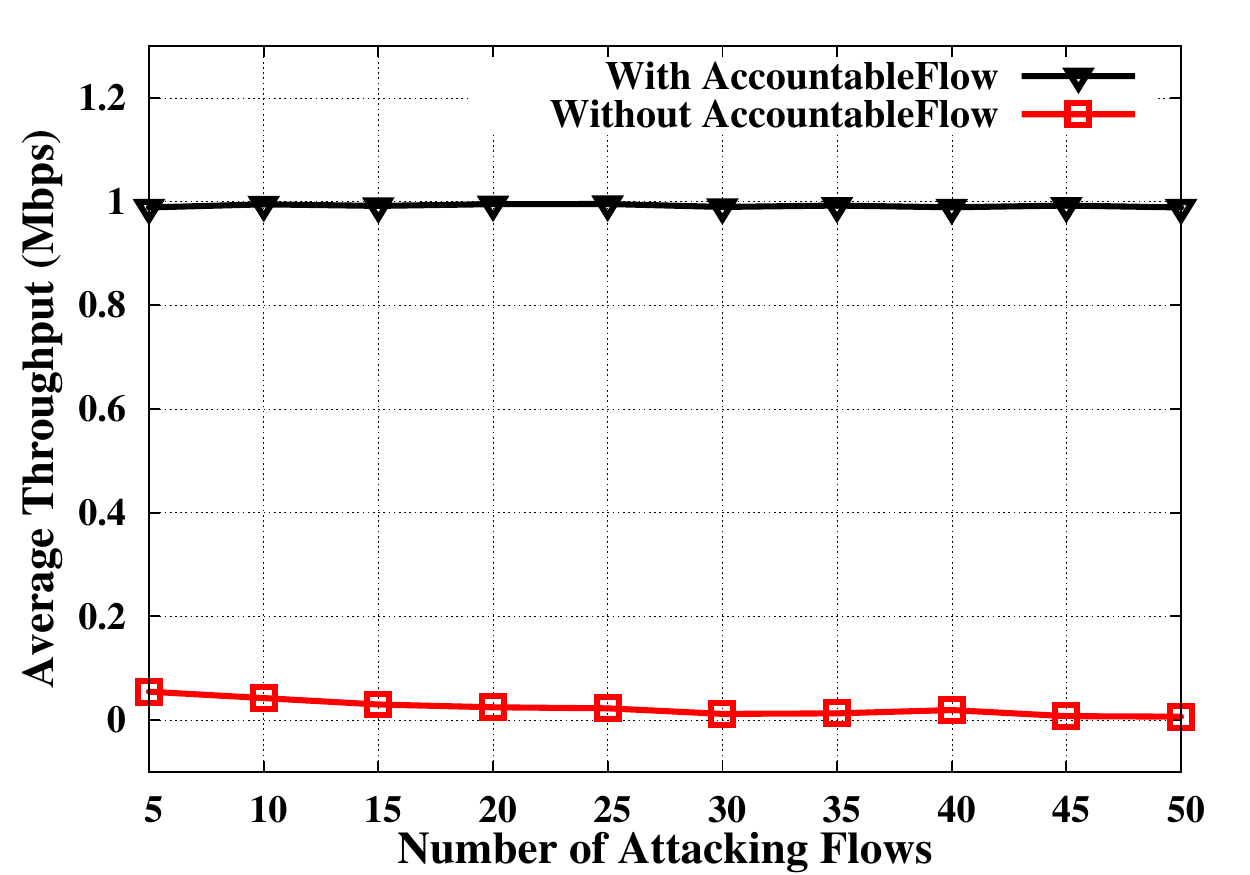}}\quad
    \subfigure[Setting Two]{\includegraphics[scale=0.46]{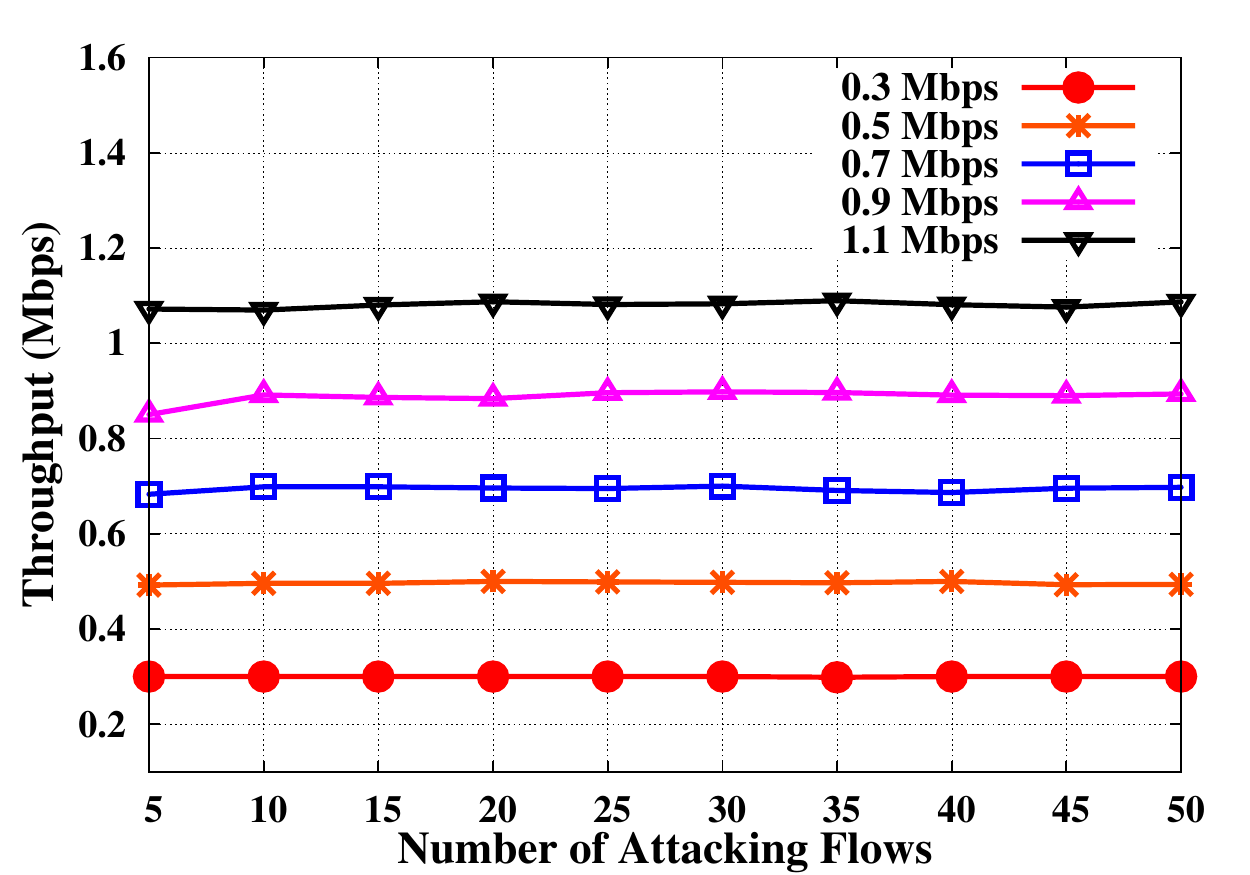}}\quad
    \subfigure[Setting Three]{\includegraphics[scale=0.46]{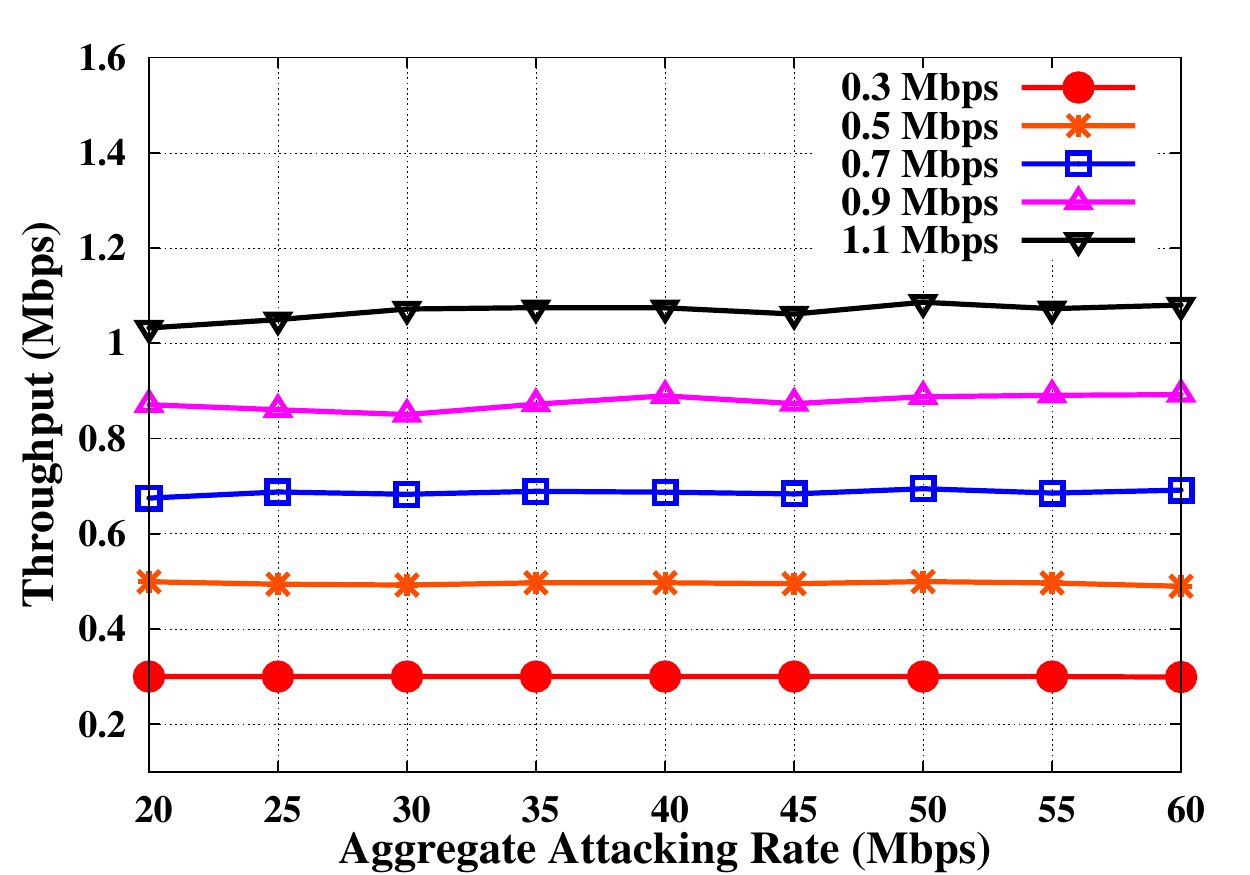}}
  }
  \caption{\sys effectively defends against the SSTF attack.}
  \label{sstfearlydrop}
\end{figure*}

We test the effectiveness of \sys when the network is faced with the SSTF attack
under similar simulation settings in Table \ref{setting}.
As illustrated in Figure \ref{sstfearlydrop}, Accountable Flow can effectively
defend against the SSTF attack. Note that we set the minimum number of
attacking flows to be $5$ since the SSTF attack needs to be distributed in order to be effective.
The convergence time is also in the order of seconds to tens of seconds.
We do not present the results for the convergence time in the paper due to space constraint.
Note that the achieved data rates (throughput) for higher rate flows,
i.e., ones with rates $0.9Mbps$ and $1.1Mbps$, are slightly less than their desired data rates.
We attribute such slight performance degradation to the fact that TCP Max-min fairness serves the low rate flows first in congested
networks.

\subsection{Benign Periodic Flow}
Real life networks, such as the Internet, also carries periodic or bursty flows whose traffic
pattern is similar to that of the low-rate TCP DoS attacking flows.
One example is that YouTube generates periodic traffic by loading
chunks of a video with pauses between each chunk \cite{youtube}.
In this subsection, we show that \sys does not falsely
drop the packets from benign periodic flows through the following four experiments.
In the first experimental setup, $5$ normal TCP flows
each transmitting at rate $1Mbps$ and one benign periodic flow
are sharing the network resources. The analysis in \cite{youtube} reveals that
the peak rate for a typical YouTube flow ranges from hundreds of kilobytes to several
megabytes. The interval of each video chunk ranges from hundreds of milliseconds to seconds. We
set the peak rate and period of the periodic flow in our simulation to be $3Mbps$ and $200ms$
so that it can represent the real video traffic.
Since the bottleneck link bandwidth is $10Mbps$ which is large enough
to hold all the traffic, the first setup is congestion free.
In the second setup, we create a fairly
congested network by generating $9$ normal TCP flows and the same periodic flow.
In the third setup, the network becomes quite congested by carrying $15$ normal TCP
flows and the same periodic flow.
Finally, in the fourth setup, the network is very congested as $20$ normal TCP flows and
the same periodic flow are traversing the bottleneck link.

\begin{figure}[ht]
\centering
\subfigure[Benign Periodic Flow.\label{benignPeriod}]{
        \begin{minipage}[t]{0.47\textwidth}
        \includegraphics[width=\textwidth]{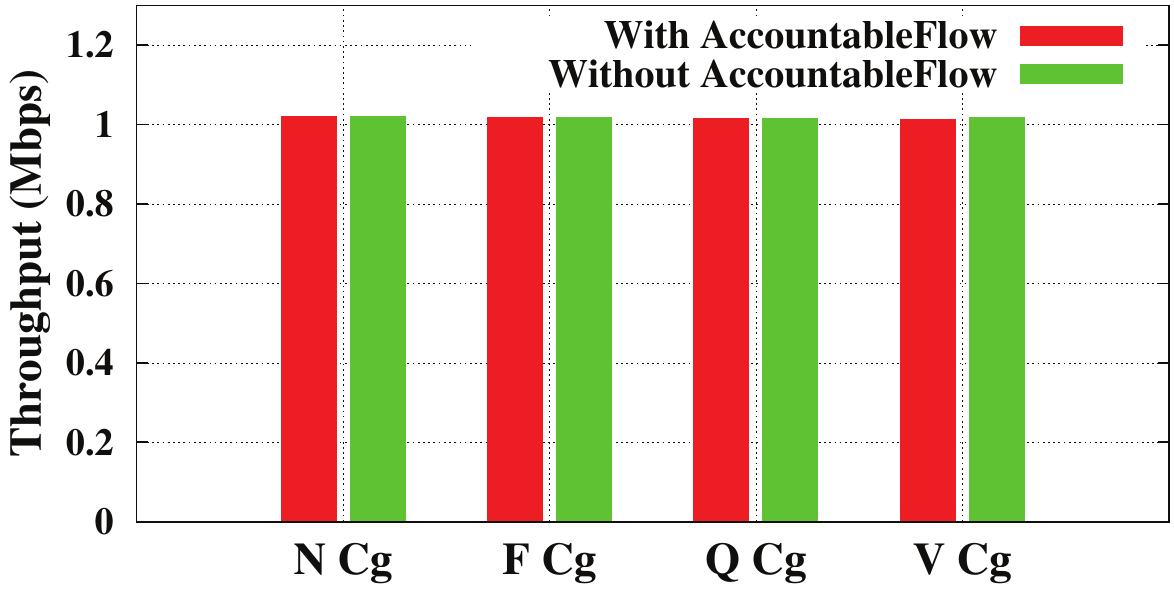}
        \end{minipage}
}
\subfigure[Normal Flows. \label{benignNormal}]{
        \begin{minipage}[t]{0.47\textwidth}
        \includegraphics[width=\textwidth]{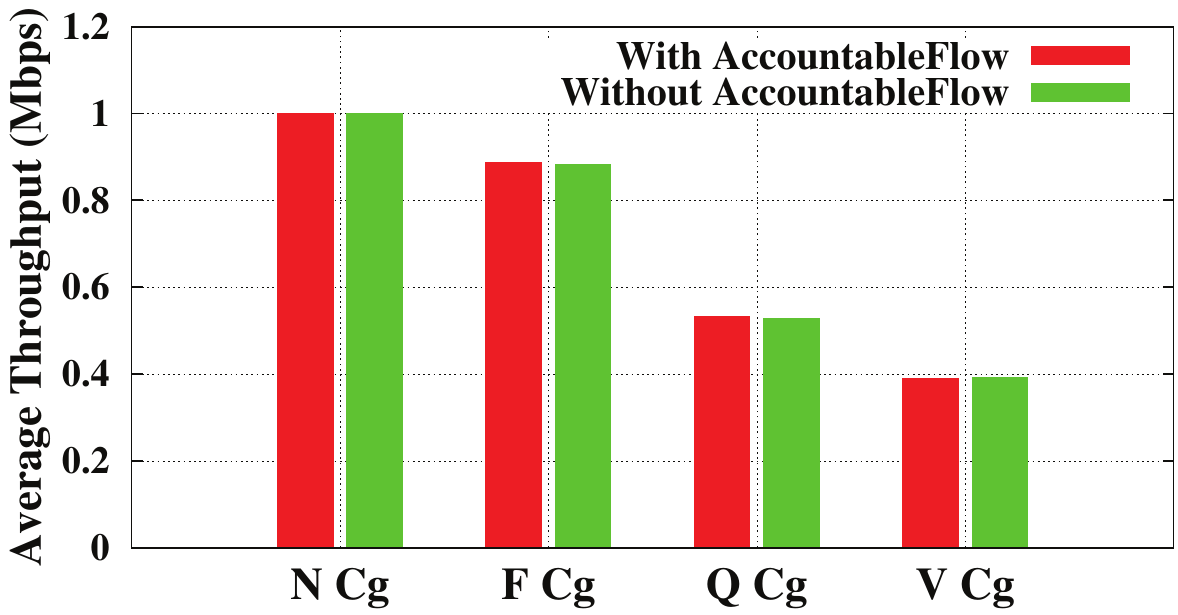}
        \end{minipage}
}
\caption{\sys harmoniously coexists with the benign periodic flow and normal flows without causing performance degradation.}\label{periodic}
\end{figure}

The simulation results are illustrated in Figure \ref{periodic}.
For clear presentation, we use characters ``N\rq\rq{}, ``F\rq\rq{}, ``Q\rq\rq{} and ``V\rq\rq{} to represent words ``Not\rq\rq{}, ``Fairly\rq\rq{},
``Quite\rq\rq{} and ``Very\rq\rq{}, respectively. The character
``Cg\rq\rq{} represents the word ``Congested\rq\rq{}. Thus
``F Cg\rq\rq{} means that the network is fairly congested, which corresponds to the second setup.
As illustrated in Figure \ref{benignPeriod}, the benign periodic flow achieves its desired throughput in all these four setups
no matter whether \sys is applied or not.
Furthermore, \sys also does not have any negative effect on the normal TCP flows, as
illustrated in Figure \ref{benignNormal}.
The results indicate that \sys can harmoniously coexist with
benign flows without causing any performance degradation.
The reason is that compared to attacking flows, the benign periodic flow is not trying to overflow the congested link.
Therefore, the network will not suffer from high aggregate loss rate. As a result, \sys
does not apply its aggressive dropping policy. Even though in some situations where a legitimate bursty flow has a
large enough  peak rate to cause
a high aggregate loss rate so that \sys drops some packets, the negative effect does not
propagate since legitimate flows will back off by entering retransmission timeouts after the packet losses. As a result, the network
will become less congested and the aggregate loss rate will drop below the threshold $Th_1$.

\begin{figure*}[t]
  \centering
  \mbox{
    \subfigure[Setting One]{\includegraphics[scale=0.46]{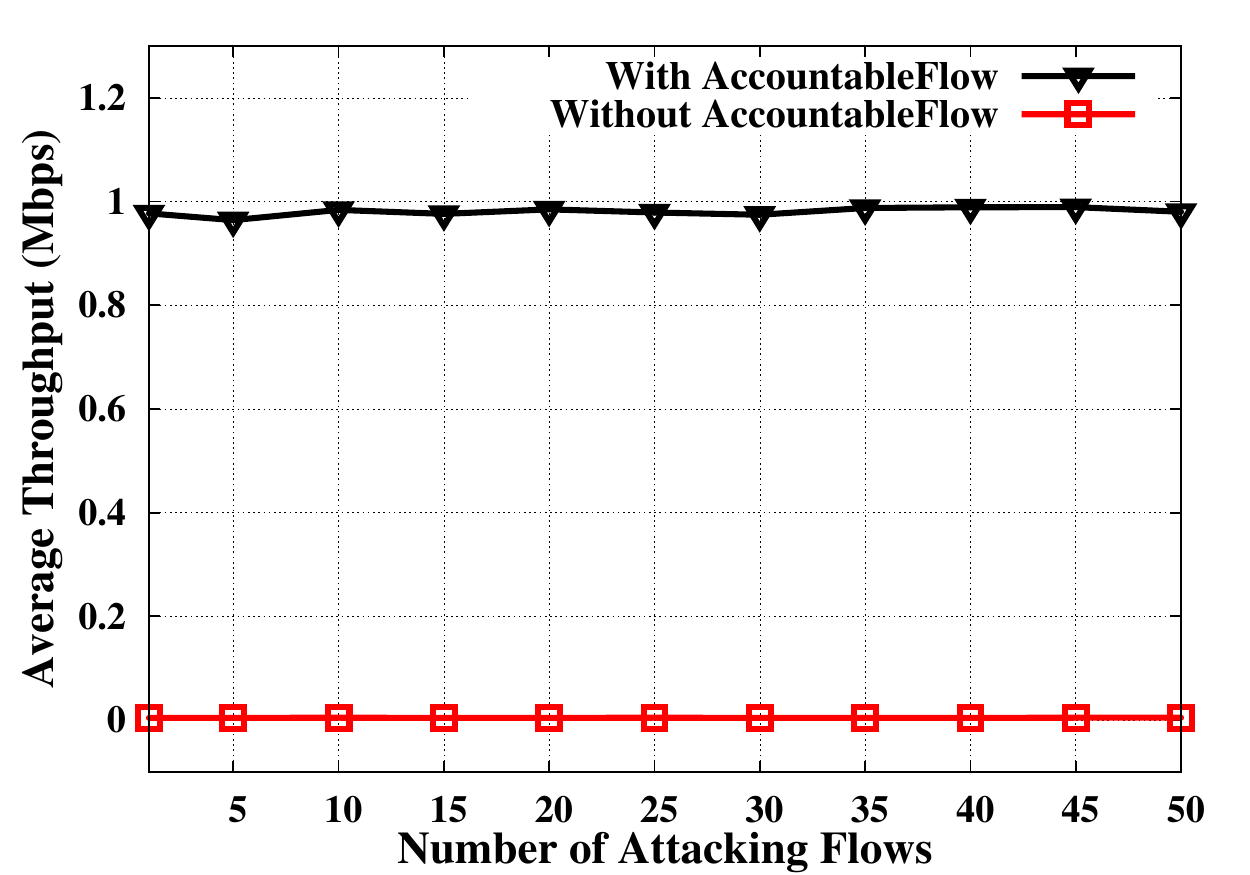}}\quad
    \subfigure[Setting Two]{\includegraphics[scale=0.46]{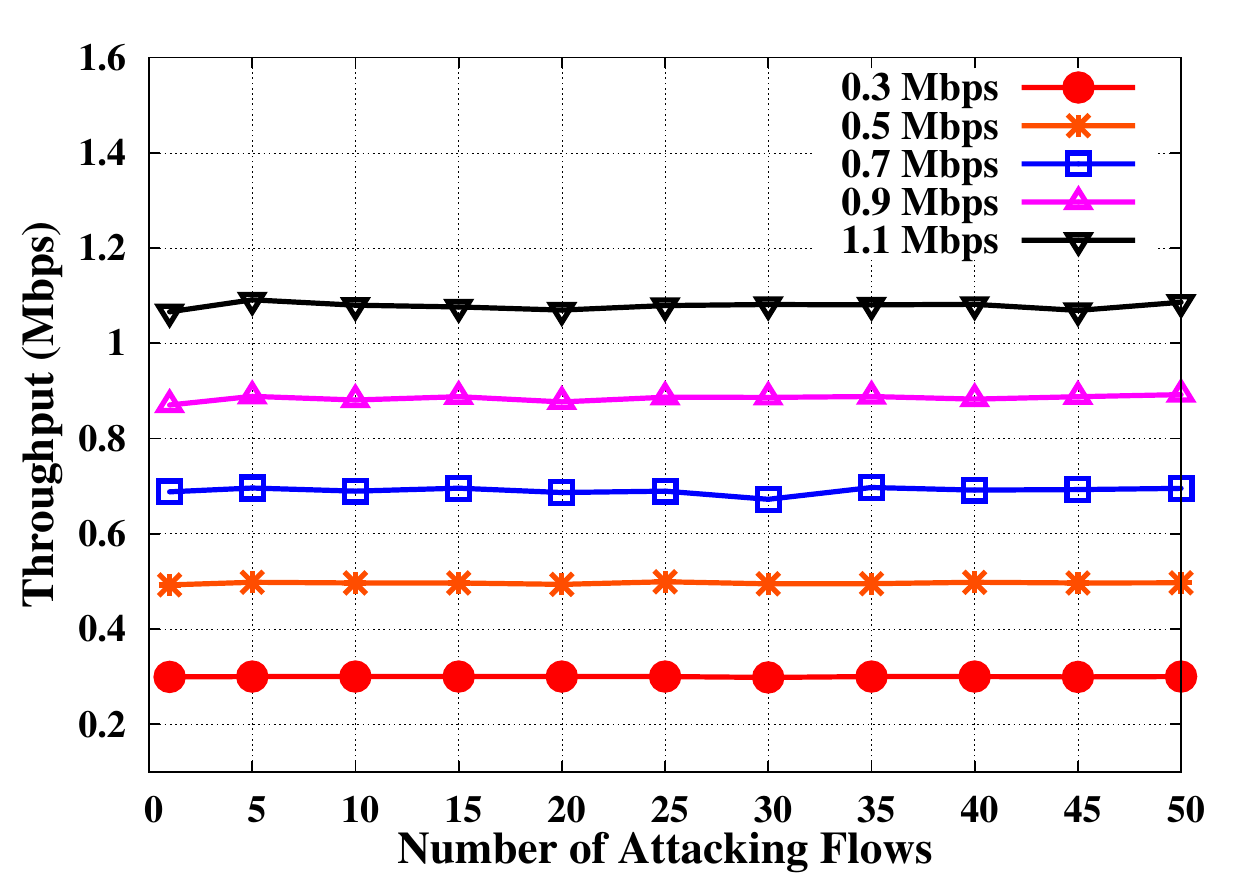}}\quad
    \subfigure[Setting Three]{\includegraphics[scale=0.46]{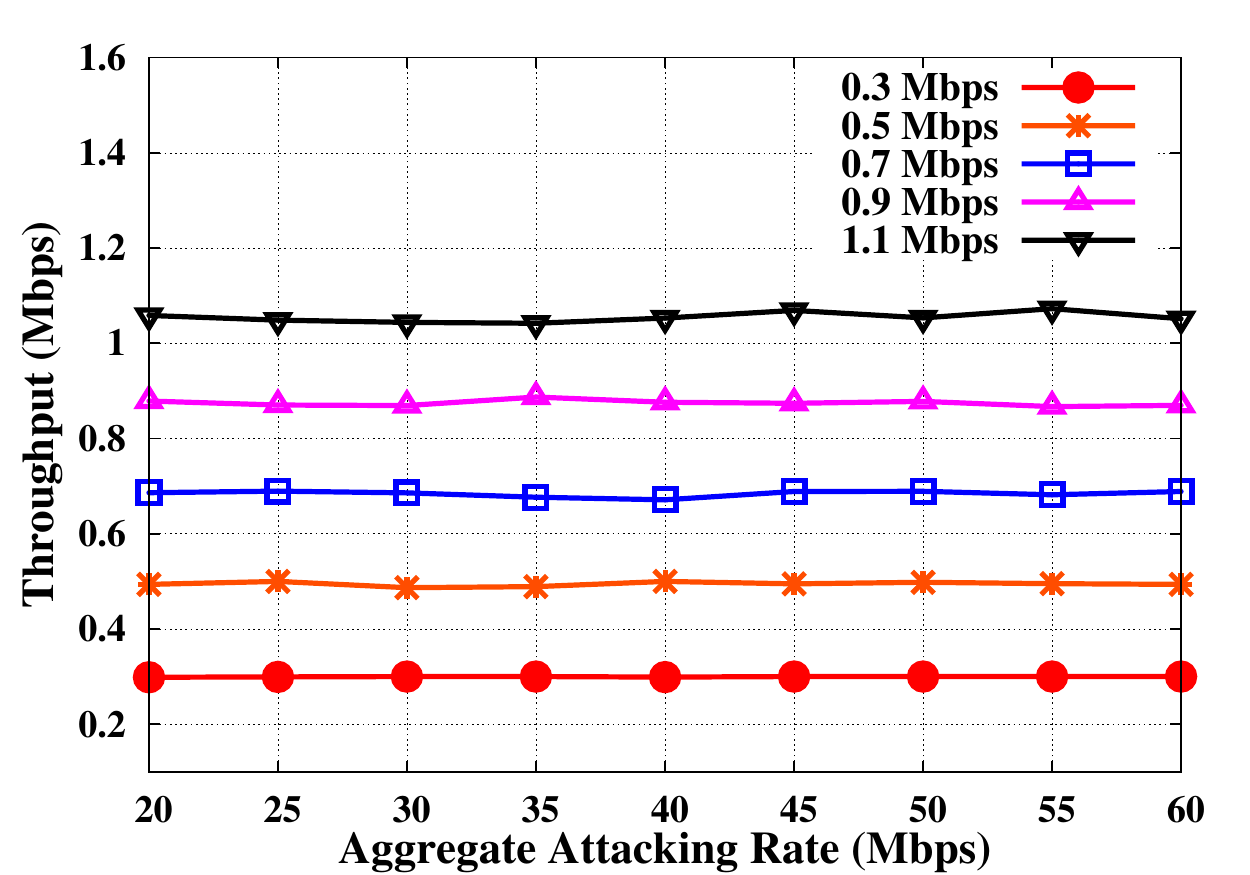}}
  }
  \caption{\sys provides a strong defense against the general DoS attacks.}
  \label{highrate}
\end{figure*}

\subsection{General (D)DoS Attack}
Although designed to solve the low-rate TCP DoS attack, \sys can also serve as the defending
technique for general DoS attacks. The difference between the low-rate
TCP DoS attack and the general DoS attacks is that the former has to rely on the TCP retransmission timeout mechanism to
launch attacks whereas the latter causes denial of service to the legitimate users simply
by sending large volumes of traffic.
As aforementioned, the core idea of \sys is to make attacking flows accountable for the
congestion by early dropping their packets according to their loss rates.
Therefore, any attacking flows that are accountable for the congestion are not able to cause
denial of service to the legitimate flows by over-utilizing network resources.
In fact, the reason why \sys is effective to deal with the general DoS
attacks is that we do not leverage on the periodic nature of the low-rate TCP DoS
attacking flows while designing the algorithm.

To test the effectiveness of \sys when the network is faced with the general DoS attacks,
we use similar simulation settings listed in Table \ref{setting}
except that the attackers consistently generate
traffic without pause. As illustrated in Figure \ref{highrate}, \sys can effectively
defend against the general DoS attacks. Furthermore, \sys also
has quick convergence time under such attacks, which is in the order of tens of seconds.

\section{Deployment of \sys}\label{other}
In this section, we consider the interaction of \sys
with other security protocols and its deployment in real networks.
Although the SDN-based security protocol has its advantages, such
as flexible control and realtime reaction to attacks,
it also has several potential problems. One of the major challenges is the scalability issue due to
the centralized network architecture. Since the centralized controller needs to create an entry
in the routing table for each distinct flow, huge numbers of distinct flows will
exhaust its the CPU and storage resources.
In this paper, we propose to use flow aggregation and virtual
centralized controller to solve the scalability problem.

\subsection{Flow Aggregation}\label{flow1}
Recall that we aggregate flows according to their source IP addresses to deal with
the SSTF attack in subsection \ref{sstfsection}. In fact, by using the aforementioned security protocols to
deal with source IP spoofing, flow aggregation can also prevent the
attackers from amplifying their attacking scale by faking huge numbers of flows.
Furthermore, with protocols like ingress filter \cite{ingress} and Passport \cite{passport},
ASes can limit the range of their acceptable source IP addresses. As a result, the
number of distinct attacking flows that can be used to attack the bottleneck link is limited. Moreover, the existing security protocols, such as MiddlePolice~\cite{middlepolice,middlepolice-tech}, Mirage \cite{DDoS1}, Phalanx \cite{million}, Pushback \cite{DDoS2} and DoS-limiting architecture \cite{DDoS3},
can be applied to further limit the attacking scales. For instance, Mirage adopts the concept of frequency hopping in wireless networks to ``hop'' the destination IP addresses among all available addresses. Each time a user wants to send traffic to this site, it has to solve a computational puzzle to get the new IP address, which will limit the volumes of traffic that the computationally limited attackers can send. MiddlePolice, on the other head, allows the destination to determine which source IPs are allowed through self-defined traffic control policies. Since \sys does not cause disruption to the existing network infrastructure, it can effectively interact with these security protocols to defend the network against extremely large scale DDoS attacks.

\subsection{Virtual Centralized Controller}\label{flow2}
\begin{figure}[h]
\begin{center}
\includegraphics[scale=0.46]{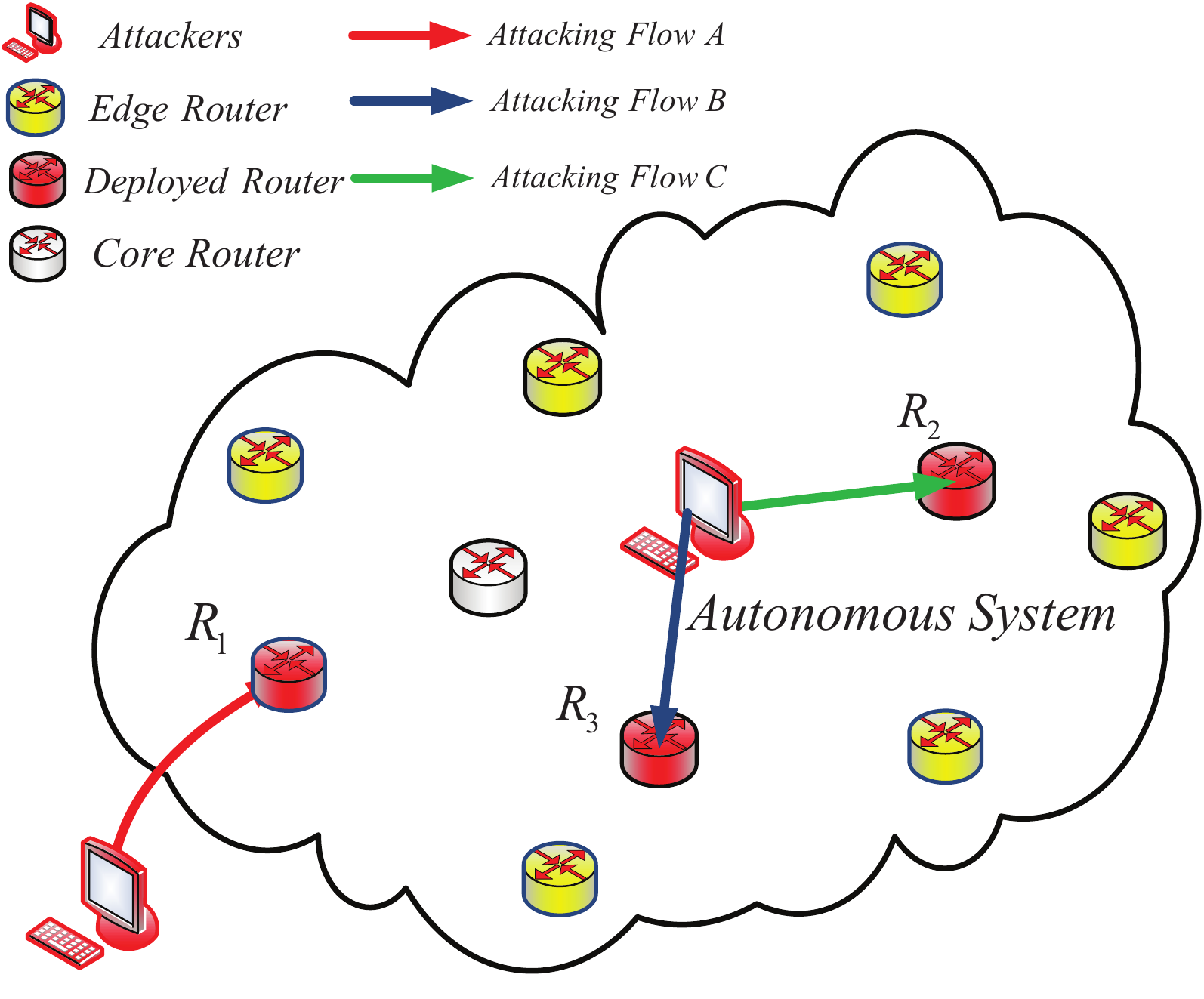}\\
\end{center}
\caption{Deployment of \sys.}\label{defendarc}
\end{figure}

In order to deal with large numbers of flows, we can also adopt the
concept of virtual centralized SDN controller. Specifically,
multiple processors can serve as the conceptual centralized controller.
Each processor keeps its routing table and tackles a certain number of flows.
In order to tolerate individual processor failures within the distributed virtual centralized controller system,
we can adopt the Paxos protocols \cite{paxos}. In fact, the B4 architecture \cite{b4},
a world wide large scale SDN data center network built by Google,
also adopts the concept of virtual centralized controller by clustering their
networks to deal with millions of flows traversing the Google's data centers.
By embracing these techniques to solve the scalability problem,
\sys can be deployed in real networks.
We present a straightforward deployment architecture in Figure \ref{defendarc}.

We consider deploying \sys on both core routers and border routers.
Typically, border routers are responsible for dealing with the inter-domain flows
such as BGP sessions \cite{bgp} whereas core routers are carrying the traffic across the
AS and may execute a particular traffic engineering policy such as MPLS \cite{mpls}.
Consider the situation where a remote legitimate client and the attacker are sending their traffic to the AS
through an undeployed border router $R_1$. Since the attacking flow $D$ exhausts the bandwidth of the
victim border router $R_1$, the client's flow $A$ is throttled to zero throughput (flow $A$ is not able to traverse $R_1$
in Figure \ref{defendarc}).
Attacking flow $D$ continues to propagate in the AS until it counters a deployed
core router $R_2$ which drops its packets to save the network resources for the legitimate flow $C$ (attacking flow
$D$ is not able to traverse $R_2$ in Figure \ref{defendarc}).
When the AS deploys \sys on its border router $R_3$, it protects the inter-domain traffic $B$
from being attacked so that $B$ can safely enter the AS.
Apart from launching inter-domain attacks, the attacker can also compromise
the nodes within the AS to generate local DoS attacking flows, such as the attacking flow $E$.
The deployed core router $R_4$ can protect the network from such an attack.
Also the deployed routers can stop the local attacking flows from leaving the AS
to attack remote sites.
To sum up, \sys is compatible with the existing security protocols and
is incrementally deployable without disruption to the existing network
infrastructure.

\section{Conclusion}\label{conclude}
In this paper we develop the \sys, an incrementally deployable SDN-based
protocol, to serve as a countermeasure against both the low-rate TCP DoS attack and general DoS attacks in WSN.
The main idea of \sys is to make the attacking flows accountable
for the congestion by dropping their packets according to their loss rates.
We test the effectiveness of \sys under four major simulation setups.
In the first setup, attackers launch low-rate TCP DoS attacks at different scales and data rates.
In the second setup, attackers vary their strategies by maliciously creating excessive numbers of
short TCP flows to occupy the network resources. The third setup is designed to
study the impact of \sys on benign flows. Finally, in the fourth setup,
attackers are launching the general DoS attacks by continuously generating traffic without
pause. Through substantial amounts of simulations in each setup, we demonstrate that
\sys, which does not cause any performance degradation to benign flows, can
effectively defend against both the low-rate TCP DoS attack and the general DoS attacks even if
attackers are able to vary their strategies.
Finally, we discuss the scalability of \sys and its deployment in real networks.


\bibliography{reference}

\begin{thebibliography}{10}
\providecommand{\url}[1]{#1}
\csname url@samestyle\endcsname
\providecommand{\newblock}{\relax}
\providecommand{\bibinfo}[2]{#2}
\providecommand{\BIBentrySTDinterwordspacing}{\spaceskip=0pt\relax}
\providecommand{\BIBentryALTinterwordstretchfactor}{4}
\providecommand{\BIBentryALTinterwordspacing}{\spaceskip=\fontdimen2\font plus
\BIBentryALTinterwordstretchfactor\fontdimen3\font minus
  \fontdimen4\font\relax}
\providecommand{\BIBforeignlanguage}[2]{{%
\expandafter\ifx\csname l@#1\endcsname\relax
\typeout{** WARNING: IEEEtran.bst: No hyphenation pattern has been}%
\typeout{** loaded for the language `#1'. Using the pattern for}%
\typeout{** the default language instead.}%
\else
\language=\csname l@#1\endcsname
\fi
#2}}
\providecommand{\BIBdecl}{\relax}
\BIBdecl

\bibitem{Siddiqui2018}
S.~Siddiqui, S.~Ghani, and A.~A. Khan, ``Adp-mac: An adaptive and dynamic
  polling-based mac protocol for wireless sensor networks,'' \emph{IEEE Sensors
  Journal}, vol.~18, no.~2, pp. 860--874, Jan 2018.

\bibitem{Zidi2018}
S.~Zidi, T.~Moulahi, and B.~Alaya, ``Fault detection in wireless sensor
  networks through svm classifier,'' \emph{IEEE Sensors Journal}, vol.~18,
  no.~1, pp. 340--347, Jan 2018.

\bibitem{Nagar2017}
S.~Nagar, S.~S. Rajput, A.~K. Gupta, and M.~C. Trivedi, ``Secure routing
  against ddos attack in wireless sensor network,'' in \emph{2017 3rd
  International Conference on Computational Intelligence Communication
  Technology (CICT)}, Feb 2017, pp. 1--6.

\bibitem{low-rate}
A.~Kuzmanovic and E.~W. Knightly, ``{Low-rate TCP-targeted denial of service
  attacks: the shrew vs. the mice and elephants},'' in \emph{ACM SIGCOMM},
  2003.

\bibitem{dash}
T.~Stockhammer, ``Dynamic adaptive streaming over http--: standards and design
  principles,'' in \emph{Proceedings of the second annual ACM conference on
  Multimedia systems}.\hskip 1em plus 0.5em minus 0.4em\relax ACM, 2011, pp.
  133--144.

\bibitem{solution1}
H.~Sun, J.~C.~S. Lui, and D.~K.~Y. Yau, ``Defending against low-rate tcp
  attacks: Dynamic detection and protection,'' in \emph{IEEE ICNP}, 2004.

\bibitem{solution2}
C.-W. Chang, S.~Lee, B.~Lin, and J.~Wang, ``The taming of the shrew: Mitigating
  low-rate tcp-targeted attack,'' \emph{IEEE TON}, 2010.

\bibitem{sdn}
N.~McKeown, T.~Anderson, H.~Balakrishnan, G.~Parulkar, L.~Peterson, J.~Rexford,
  S.~Shenker, and J.~Turner, ``Openflow: Enabling innovation in campus
  networks,'' \emph{ACM SIGCOMM}, 2008.

\bibitem{b4}
S.~Jain, A.~Kumar, S.~Mandal, J.~Ong, L.~Poutievski, A.~Singh, S.~Venkata,
  J.~Wanderer, J.~Zhou, M.~Zhu, J.~Zolla, U.~H\"{o}lzle, S.~Stuart, and
  A.~Vahdat, ``B4: Experience with a globally-deployed software defined wan,''
  \emph{ACM SIGCOMM}, 2013.

\bibitem{swan}
C.-Y. Hong, S.~Kandula, R.~Mahajan, M.~Zhang, V.~Gill, M.~Nanduri, and
  R.~Wattenhofer, ``Achieving high utilization with software-driven wan,'' in
  \emph{ACM SIGCOMM}, 2013.

\bibitem{sdns}
S.~Shin, P.~Porras, V.~Yegneswaran, M.~Fong, G.~Gu, and M.~Tyson, ``Fresco:
  Modular composable security services for software-defined networks,'' in
  \emph{Proceedings of Network and Distributed Security Symposium}, 2013.

\bibitem{flowpolice}
Z.~Liu, ``{FlowPolice: Enforcing Congestion Accountability to Defend against
  DDoS Attacks},'' Ph.D. dissertation, University of Illinois at
  Urbana-Champaign, 2015.

\bibitem{ns3}
\BIBentryALTinterwordspacing
``Ns-3: a discrete-event network simulator.'' [Online]. Available:
  \url{http://www.nsnam.org/}
\BIBentrySTDinterwordspacing

\bibitem{maxmin}
E.~L. Hahne, ``Round-robin scheduling for max-min fairness in data networks,''
  \emph{IEEE J.Sel. A. Commun.}

\bibitem{ip1}
A.~Yaar, A.~Perrig, and D.~Song, ``Stackpi: New packet marking and filtering
  mechanisms for ddos and ip spoofing defense,'' \emph{Selected Areas in
  Communications, IEEE Journal on}, vol.~24, no.~10, pp. 1853--1863, 2006.

\bibitem{ip2}
Z.~Duan, X.~Yuan, and J.~Chandrashekar, ``Controlling ip spoofing through
  interdomain packet filters,'' \emph{Dependable and Secure Computing, IEEE
  Transactions on}, vol.~5, no.~1, pp. 22--36, 2008.

\bibitem{youtube}
P.~Ameigeiras, J.~J. Ramos-Munoz, J.~Navarro-Ortiz, and J.~M. Lopez-Soler,
  ``Analysis and modelling of youtube traffic,'' \emph{Transactions on Emerging
  Telecommunications Technologies}, vol.~23, no.~4, pp. 360--377, 2012.

\bibitem{ingress}
P.~Ferguson and D.~Senie, ``Network ingress filtering: Defeating denial of
  service attacks which employ ip source address spoofing,'' United States,
  2000.

\bibitem{passport}
X.~Liu, A.~Li, X.~Yang, and D.~Wetherall, ``Passport: Secure and adoptable
  source authentication,'' in \emph{USENIX NSDI}, 2008.

\bibitem{middlepolice}
Z.~Liu, H.~Jin, Y.-C. Hu, and M.~Bailey, ``{MiddlePolice: Toward Enforcing
  Destination-Defined Policies in the Middle of the Internet},'' in \emph{ACM
  CCS}, 2016.

\bibitem{middlepolice-tech}
{Z. Liu, H. Jin, Y.-C. Hu, M. Bailey}, ``{MiddlePolice: Fine-Grained
  Endpoint-Driven In-Network Traffic Control for Proactive DDoS Attack
  Mitigation},'' 2017.

\bibitem{DDoS1}
P.~Mittal, D.~Kim, Y.-C. Hu, and M.~Caesar, ``Mirage: Towards deployable ddos
  defense for web applications,'' \emph{arXiv preprint arXiv:1110.1060}, 2011.

\bibitem{million}
C.~Dixon, T.~Anderson, and A.~Krishnamurthy, ``Phalanx: Withstanding
  multimillion-node botnets,'' in \emph{USENIX NSDI}, 2008.

\bibitem{DDoS2}
R.~Mahajan, S.~M. Bellovin, S.~Floyd, J.~Ioannidis, V.~Paxson, and S.~Shenker,
  ``Controlling high bandwidth aggregates in the network,'' \emph{ACM SIGCOMM},
  2002.

\bibitem{DDoS3}
X.~Yang, D.~Wetherall, and T.~Anderson, ``A dos-limiting network
  architecture,'' in \emph{ACM SIGCOMM}, 2005.

\bibitem{paxos}
T.~Chandra, R.~Griesemer, and J.~Redstone, ``Paxos made live-an engineering
  perspective (2006 invited talk),'' in \emph{Proceedings of the 26th ACM
  Symposium on Principles of Distributed Computing-PODC}, vol.~7, 2007.

\bibitem{bgp}
S.~Kent, C.~Lynn, and K.~Seo, ``Secure border gateway protocol (s-bgp),''
  \emph{IEEE JSAC}, 2000.

\bibitem{mpls}
D.~O. Awduche and J.~Agogbua, ``Requirements for traffic engineering over
  mpls,'' 1999.

\end{thebibliography}
\bibliographystyle{IEEEtran}


\end{document}